# Analytical Pricing of Defaultable Bond with Stochastic Default Intensity

– The Case with Exogenous Default Recovery –

**O, Hyong-chol**[1,2]     **Wan, Ning**[2]

1) Centre of Basic Sciences, **Kim Il Sung** University, Pyongyang, D. P. R. of Korea,
2) Department of Applied Mathematics, Tong-ji University, Shanghai, China
Authors: O Hyongchol, 1964-, Researcher; Wan Ning, 1979-, Graduate Student
(Corresponding Author: Tel: +86-021-65988317; Fax: +86+021-65982342, E-mail: blumping@etang.com
Mailing Address:   Department of Applied Mathematics, Tong-ji University, Shanghai, 200092, China,
….

**Abstract:** Here we provide analytical pricing formula of corporate defaultable bond with both expected and unexpected default in the case with stochastic default intensity. In the case with constant short rate and exogenous default recovery using PDE method, we gave some pricing formula of the defaultable bond under the conditions that 1)expected default recovery is the same with unexpected default recovery; 2) default intensity follows one of 3 special cases of Willmott model; 3) default intensity is uncorrelated with firm value. Then we derived a price formula of a credit default swap. And in the case of stochastic short rate and exogenous default recovery using PDE method, we gave some pricing formula of the defaultable bond under the conditions that 1) expected default recovery is the same with unexpected default recovery; 2) the short rate follows Vasicek model; 3) default intensity follows one of 3 special cases of Willmott model; 4) default intensity is uncorrelated with firm value; 5) default intensity is uncorrelated with short rate. Then we derived a price formula of a credit default swap. We give some credit spread analysis, too.

## $ 1. Introductions

There are two main approaches to pricing defaultable corporate bonds; one is the *structural approach* and the other one is the *reduced form approach*.

The structural approach is initiated by Merton, Black and Cox, Geske, and subsequently developed by Longstaff and Schwartz and etc. In the structural method, we think that the default event occurs when the firm value reaches a certain lower threshold (*default barrier*) from the above. Such a default can be expected. According to the structural approach's idea, if the firm value is larger than the default threshold very much, then we could think the firm has no default risk. But in reality, the corporate bond price is also cheaper than the default free zero coupon bond price in such a circumstance. This difference could be seen as a reflection of the unexpected default risk of corporate bonds, so structural approach cannot reflect all of credit risks, it means that the structural approach *underestimates* credit risk. This is the shortcoming of the structural approach.

In the reduced-form approach, the default is treated as an unpredictable event governed by a default intensity process, the default event has no any correlation with the firm value. So the default might not occur in very low asset values and therefore default event would be unexpected. Duffie - Singleton, Jarrow - Turnbull, Jarrow-Lando-Turnbull and some other people studied the reduced form model. In the reduced-form approach, the default probability in infinitesimal time interval is called a default intensity and the difference of corporate bond price from the default free zero coupon bond price is considered to come from credit risk, and default intensity is defined using this difference. But in fact, the difference of corporate bond price from the default free zero coupon bond price could depend not only on credit risk but also on other factors, thus the reduced form model *overrates* credit risk. This is the shortcoming of the reduced form approach.

In order to overcome the shortcoming of every approach and improve the quality of modeling, one of the recent trends is to combine elements of the two approaches. [Belanger-Shreve-Wong 2001, Cathcart-El Jahel 2003, Realdon 2004]

Cathcart-El Jahel (2003) using PDE method provided a semi-analytical pricing formula of defaultable bond with both expected and unexpected default when the short rate follows CIR model and default intensity is linearly dependent on the short rate. They studied the case with exogenous default recovery and no correlation of firm value and short rate, and they assumed that unexpected default recovery and expected default recovery are the same.

Jiang-Luo (2005) studied the same problem with Cathcart-El Jahel (2003), but in their model, short rate follows Vasicek model and they assume that correlation of firm value and short rate is not necessarily zero, and furthermore they provided analytical pricing formula using PDE method.

Realdon(2004) using PDE method provided an analytical pricing formula of defaultable bond with both expected and unexpected default in the case with constant short rate and endogenous default recovery when the default intensity is constant. And they also provided an analytical pricing formula in the case that short rate is uncorrelated with firm value and default recovery is exogenous when the default intensity is constant. They didn't assume that unexpected and expected default recoveries are the same.

In this paper, we study the pricing problem of defaultable bond with both expected and unexpected default in the case with *stochastic default intensity*.

Wilmott (1998) established a very general model for default intensity (hazard rate) process and using PDE method provided some pricing formula of defaultable bond with unexpected default only and no default recovery. On the other hand, from investment grade corporate bonds data, Collin-Solnik (2001) obtain maximum likelihood estimates of the parameters when unexpected default intensity follows Vasicek-like model (Wilmott model includes it). This fact supports the usefulness of Wilmott model for default intensity in real data.

In this paper, we provide an analytical pricing formula of corporate defaultable bond with both expected and unexpected default in the case with stochastic default intensity. In the case with constant short rate and exogenous default recovery using PDE method, we gave some pricing formula of the defaultable bond under the conditions that 1) expected default recovery is the same with unexpected default recovery; 2) default intensity follows one of 3 special cases of Wilmott model; 3) default intensity is uncorrelated with firm value. And then we derived a pricing formula of a credit default swap. In the case with stochastic short rate (three factor model) and exogenous default recovery using PDE method, we gave some pricing formula of the defaultable bond under the conditions that 1) expected default recovery is the same with unexpected default recovery; 2) the short rate follows Vasicek model; 3) default intensity follows one of 3 special cases of Willmott model; 4) default intensity is uncorrelated with firm value; 5) default intensity is uncorrelated with short rate. And then we derived a pricing formula of a credit default swap. We gave some credit spread analysis.

The remainder of this paper is organized as follows: Section 2 and 3 deal with constant short rate and section 4 and 5 deal with stochastic short rate.

In section 2, we studied the pricing problem of corporate zero coupon bond with unexpected default only. Its aim is to clarify the effect of stochastic unexpected default intensity. Here we assume that default intensity process follows Wilmott model and establish a pricing PDE model for corporate zero coupon bond with unexpected default only, and then using the structural solution method, we provide no default probability formulae and pricing formulae of defaultable bonds with face value exogenous recovery and market price exogenous recovery in the 3 special cases of Wilmott model for default intensity.

In section 3 we studied the pricing formula of the corporate zero coupon bond with both of expected default and unexpected default. Here we used the "Corporate Zero Coupon Bond with Unexpected Default only" (considered in previous section) as a "*virtual bond*" *to hedge the risk* of unexpected default in establishing the pricing PDE model. And then using the separating variables method, we provide no default probability formula and pricing formula of defaultable bond with face value exogenous recovery in the 3 special cases of Wilmott model for default intensity. In the end of the section we derived a price formula of a credit default swap.

In section 4, we studied the pricing problem of corporate zero coupon bond with unexpected default only. Its aim is to clarify the effect of stochastic unexpected default intensity. Here we assume that the short rate follows Vasicek model and default intensity process follows Wilmott model and establish a pricing PDE model for corporate zero coupon bond with unexpected default only. And then using the method of separating variables and the result of section 2, we provide no default probability formula and pricing formula of defaultable bond with face value exogenous recovery in the 3 special cases of Wilmott model for default intensity. Next using the structural solution method, we provide no default probability formula and pricing formula of defaultable bond with market price exogenous recovery in another 3 special cases (more general than before) of Wilmott model for default intensity.

In section 5 we studied the pricing formula of the corporate zero coupon bond with both of expected default and unexpected default. Here as in section 3, we used the "Corporate Zero Coupon Bond with Unexpected Default only" (considered in previous section) as a "*virtual bond*" *to hedge the risk* of unexpected default in establishing the pricing PDE model. And then using the change of numeraire by default free zero coupon bond and the method of separating variables, we provide no default probability formula and pricing formula of defaultable bond with face value exogenous recovery in the 3 special cases of Wilmott model for default intensity. In the end of the section we derived a price formula of a credit default swap.

In section 6, we give some credit spread analysis.

## $ 2. The Pricing Corporate Zero Coupon Bond with Unexpected Default only.

In order to clarify the effect of stochastic unexpected default intensity, we consider the simplest case that the price of corporate zero coupon bond is not effected by the firm value but only effected by unexpected default.

Main assumptions are as follows:

**Assumption** 1: The short-term interest rate $r$ is constant.

**Assumption** 2: The unexpected default probability in $[t, t + dt]$ is $\boldsymbol{p_t} dt$, the default intensity $\boldsymbol{p_t}$ follows

$$dp = a(p,t)dt + s(p,t)dW \quad \text{(on natural measure)} \tag{2.1}$$

**Assumption** 3: Default recovery is given as the form of face value exogenous recovery ($R \cdot e^{-r(T-t)}$, $R$: const) or as the form of market price exogenous recovery ($R \times$ bond price at default time).

**Assumption** 4: Our Corporate bond price is given by the function $\hat{C} = \hat{C}(p,t)$.

**A Financial explanation about assumptions**:

We suppose that such a model is not only useful when we do not know the circumstance of the firm value but also used as a "*virtual bond*" *to hedge the risk* of unexpected default even when we know very well about the circumstance of the firm value, as shown in next section. In assumption 3 face value exogenous recovery means that default recovery is the same with paying **R** at maturity.

**The Deriving of PDE model.**

Here, we have no any underlying asset with which to hedge the risk of $p_t$. The only traded asset is our bond as a derivative of our independent variable $p_t$ and so the only way to construct a hedged portfolio is by hedging one bond with a bond of a different maturity [Wilmott 98].

Let denote the price of a bond with maturity $T_i$ and default recovery $R_i$ by $C_i(p, t; T_i)$, i = 1, 2.

Construct a portfolio:
$$\Pi = C_1 - \Delta C_2$$
such that it is risk-free, that is,
$$d\Pi = dC_1 - \Delta dC_2 = r\Pi dt \tag{2.2}$$

(i) The case of no default (with probability $1 - p_t dt$)

$$d\Pi = \left(\frac{\partial C_1}{\partial t} + \frac{1}{2}s^2 \frac{\partial^2 C_1}{\partial p^2}\right)dt + \frac{\partial C_1}{\partial p}dp - \Delta\left[\left(\frac{\partial C_2}{\partial t} + \frac{1}{2}s^2 \frac{\partial^2 C_2}{\partial p^2}\right)dt + \frac{\partial C_2}{\partial p}dp\right]$$

So let

$$\Delta = \frac{\partial C_1}{\partial p}\left(\frac{\partial C_2}{\partial p}\right)^{-1}, \tag{2.3}$$

then

$$d\Pi = \left[\frac{\partial C_1}{\partial t} + \frac{1}{2}s^2 \frac{\partial^2 C_1}{\partial p^2} - \frac{\partial C_1}{\partial p}\left(\frac{\partial C_2}{\partial p}\right)^{-1}\left(\frac{\partial C_2}{\partial t} + \frac{1}{2}s^2 \frac{\partial^2 C_2}{\partial p^2}\right)\right]dt, \tag{2.4}$$

(ii) The case of default (with probability $p_t dt$)

$$d\Pi = (R_1 - C_1) - \frac{\partial C_1}{\partial p}\left(\frac{\partial C_2}{\partial p}\right)^{-1}(R_2 - C_2), \tag{2.5}$$

(2.4)$\times$ ($1 - p_t dt$) + (2.5)$\times p_t dt$, considering (2.2) and neglecting the higher order infinitesimal of **dt**, then we have

$$\frac{\frac{\partial C_1}{\partial t} + \frac{1}{2}s^2 \frac{\partial^2 C_1}{\partial p^2} - (r+p)C_1 + R_1 p}{\frac{\partial C_1}{\partial p}} = \frac{\frac{\partial C_2}{\partial t} + \frac{1}{2}s^2 \frac{\partial^2 C_2}{\partial p^2} - (r+p)C_2 + R_2 p}{\frac{\partial C_2}{\partial p}}.$$

The left hand of this equation is a function of $T_1$ but not $T_2$, and the right hand is a a function of $T_2$ but not $T_1$. Thus the both sides must be a function – $a_p(p, t)$ independent of maturity date. So we have the equation of corporate bond with unexpected default intensity $p_t$:

$$\frac{\partial \hat{C}}{\partial t} + \frac{1}{2}s^2 \frac{\partial^2 \hat{C}}{\partial p^2} + a_p(p,t)\frac{\partial \hat{C}}{\partial p} - (r+p)\hat{C} + Rp = 0. \tag{2.6}$$

Denote by

$$\lambda(p,t) = \frac{a(p,t) - a_p(p,t)}{s(p,t)}, \text{ that is, } a_p(p,t) = a(p,t) - s(p,t)\lambda(p,t), \tag{2.7}$$

$\lambda(p, t)$ is called a **market price of risk** $p_t$ and $a_p(p, t)$ is called a **risk neutral drift** of $p_t$, that is, in the risk neutral measure, (2.1) is changed to

$$dp = a_p(p,t)dt + s_p(p,t)dW_1, \text{ (on risk neutral martingale measure)} \tag{2.8}$$

here $s_p(p, t) = s(p, t)$ is not changed.

Thus remove the mountain symbol ^ for simplicity and consider the recovery assumption 3, then we have the *PDE model for pricing corporate bond with unexpected default intensity $p_t$ only*:

$$\begin{cases} \frac{\partial C}{\partial t} + \frac{1}{2}s_p^2(p,t)\frac{\partial^2 C}{\partial p^2} + a_p(p,t)\frac{\partial C}{\partial p} - (r+p)C + R \cdot e^{-r(T-t)} p = 0, (0 < t < T, p > 0) \\ C(p,T) = 1, \quad (p > 0). \end{cases} \tag{2.9}$$

It is a terminal value problem of *non homogeneous* parabolic equation with *variable coefficients*. Use numeraire as bank save account, that is,

$$C = ue^{-r(T-t)}, \tag{2.10}$$

Then we get the *non homogeneous* equation with *risk free rate* **0**:

$$\begin{cases} \frac{\partial u}{\partial t} + \frac{1}{2}s_p^2(p,t)\frac{\partial^2 u}{\partial p^2} + a_p(p,t)\frac{\partial u}{\partial p} - p(u - R) = 0, (0 < t < T, p > 0) \\ u(p,T) = 1, \quad (p > 0). \end{cases}$$

Use change of unknown function

$$\hat{u} = u - R, \tag{2.11}$$

Then we get the *homogeneous* equation with *risk free rate* **0** ;

$$\begin{cases} \frac{\partial \hat{u}}{\partial t} + \frac{1}{2}s_p^2(p,t)\frac{\partial^2 \hat{u}}{\partial p^2} + a_p(p,t)\frac{\partial \hat{u}}{\partial p} - p\hat{u} = 0, (0 < t < T, p > 0) \\ \hat{u}(p,T) = 1 - R, \quad (p > 0). \end{cases} \tag{2.12}$$

Use again change of unknown function

$$\hat{u} = W(1 - R), \tag{2.13}$$

Then we get

$$\begin{cases} \dfrac{\partial W}{\partial t} + \dfrac{1}{2} s_p^{\,2}(p,t) \dfrac{\partial^2 W}{\partial p^2} + a_p(p,t) \dfrac{\partial W}{\partial p} - pW = 0, (0 < t < T, p > 0) \\ W(p,T) = 1, \quad (p > 0). \end{cases} \tag{2.14}$$

From (2.10), (2.11) and (2.13), our defaultable bond price is represented by $W$ as

$$\begin{aligned} C(p,t) &= ue^{-r(T-t)} = (\hat{u} + R)e^{-r(T-t)} = [W(1-R) + R]e^{-r(T-t)} \\ &= W(p,t)e^{-r(T-t)} + (1-W)R \cdot e^{-r(T-t)} \end{aligned} \tag{2.15}$$

That is, the price of our defaultable bond at time $t$ can be seen as an expectation of the current value of the bond in the case of no default and the value of the bond in the case of default at time $t$. So $W(p, t)$ in (2.13) can be regarded as *a probability at time $t$ of no default* in the case of default intensity $p$ and thus the equation (2.14) is a **PDE equation of no default probability** when default intensity is $p$.

In order to solve (2.14), we **restrict** ourselves into the following case as in [**Wilmott**:1998]:

**Assumption** 5): In (2.8), $a_p(p, t)$ and $s_p^2(p, t)$ are linear on $p$, that is

$$a_p(p,t) = b(t) - c(t)p; \quad s_p^{\,2}(p,t) = d(t) + e(t)p. \tag{2.16}$$

Then we have the solution of (15) in the form of

$$W(p,t) = e^{A(t,T) - B(t,T)p}. \tag{2.17}$$

We can find the explicit expressions of $A(t, T)$ and $B(t, T)$. (see appendix 1.)

$$B(t,T) = \begin{cases} \dfrac{1-\exp(-c(T-t))}{c}, & dp_t = (b(t) - c \cdot p)dt + \sqrt{d(t)} \cdot dW, c \neq 0, \\ T-t, & dp_t = b(t)dt + \sqrt{d(t)} \cdot dW, \\ \sqrt{2/C}\,\text{th}[\sqrt{C/2}(T-t)], & dp_t = b(t)dt + \sqrt{d(t) + C \cdot p} \cdot dW, C > 0. \end{cases} \tag{2.18}$$

$$A(t,T) = -\int_t^T [b(s)B(s) - \frac{1}{2}d(s)B^2(s)]ds \tag{2.19}$$

Thus, In the case when the default intensity $p_t$ satisfies one of 3 models in (2.18), the **price of corporate bond with unexpected default** intensity $p_t$ is given by

$$C(p,t) = e^{A(t,T) - B(t,T)p} e^{-r(T-t)} + (1 - e^{A(t,T) - B(t,T)p})R \cdot e^{-r(T-t)}, \tag{2.20}$$

$$= e^{-r(T-t)}R + e^{-r(T-t)}(1-R)e^{A(t,T) - B(t,T)p}, \tag{2.21}$$

where $B(t, T)$ and $A(t, T)$ are respectively given by (2.18) and (2.19).

**Remark**: The formulae (2.20) and (2.21) are also true in the case when the default intensity $p_t$ satisfies the models (2.16) but do not satisfy one of the conditions of (2.18), although in this case $B(t, T)$ cannot have an explicit formula.

The *financial meaning* of (2.20) is clear as mentioned in the above. The *financial meaning* of (2.21) is as follows: the *first term* of (2.21) is the current price of *the part to be given to bond holder regardless of default occurs or not*, and the *second term* is the allowance dependent on default probability at time *t*. If at some moment *t*, the default is certain ($W(p, t) = 0$), then the price of the bond at *t* is exactly the current price of default recovery. And as shown in formula (2.21), if $R = 0$, that is, if there is nothing to recover in case of default, then the bond price is the product of zero coupon bond and survival probability; and if $R = 1$, that is, if there is no any loss in case of default, then the bond price is the same with zero coupon bond.

*Note*: The case of market price recovery : Then the pricing equation (2.9) becomes

$$\begin{cases} \dfrac{\partial C}{\partial t} + \dfrac{1}{2} s_p^2(p,t) \dfrac{\partial^2 C}{\partial p^2} + a_p(p,t) \dfrac{\partial C}{\partial p} - (r + p(1-R))C = 0, (0 < t < T, p > 0) \\ C(p,T) = 1, \quad (p > 0). \end{cases} \quad (2.22)$$

It is a terminal value problem of **homogeneous** parabolic equation with **variable coefficients**. Use numeraire as bank save account, that is,

$$C = u e^{-r(T-t)}$$

Then we get the homogeneous equation with **risk free rate 0**:

$$\begin{cases} \dfrac{\partial u}{\partial t} + \dfrac{1}{2} s_p^2(p,t) \dfrac{\partial^2 u}{\partial p^2} + a_p(p,t) \dfrac{\partial u}{\partial p} - p(1-R)u = 0, (0 < t < T, p > 0) \\ u(p,T) = 1, \quad (p > 0). \end{cases} \quad (2.23)$$

This equation has the same type with (2.14), so under the assumption (2.16), it has the solution of following type:

$$u(p,t) = e^{A(t,T) - B(t,T)p}. \quad (2.24)$$

Here *A*, *B* are similarly given as in the above.

## $ 3. The Pricing Corporate Zero Coupon Bond with Default Barrier.

In this section, we studied the pricing formula of the Corporate Zero Coupon Bond with both of expected default and unexpected default. Here we used the "Corporate Zero Coupon Bond with Unexpected Default only" (considered in previous section) as a "*virtual bond*" to hedge the risk of unexpected default in deriving the pricing PDE.

Main Assumptions are as follows:

**Assumption** 1: Short-term rate *r* is constant;

**Assumption** 2: The firm value *V* consists of *m* shares of *traded* assets *S* and *n* sheets of *zero coupon bonds*:

$$V_t = mS_t + nC_t \quad (3.1)$$

and follows the geometric Brownian motion, (drift $a_V$, volatility $s_V$: constants)

$$dV = a_V V dt + s_V V dW_2, \quad (\text{natural measure}), \quad (3.2)$$

The firm pays out dividend in ratio *b* (const) for a unit of firm value, continuously.

**Assumption** 3: Unexpected default probability in [$t$, $t + dt$] is $p_t dt$, and the default intensity $p_t$ satisfies (2.1);

**Assumption** 4: Expected default occurs when

$$V \leq V_b(t); \quad V_b(t) = V_B \text{ or } V_B e^{-r(T-t)}, V_B : \text{const.}$$

**Assumption** 5: Expected and unexpected default recovery is $R_d = R \cdot e^{-r(T-t)}; 0 \leq R \leq 1 : \text{const}$

**Assumption** 6: $E(dW_1 \cdot dW_2) = 0$, that is, *unexpected default is **not correlated*** with firm value. (this is a very likable assumption);

**Assumption** 7: Our Corporate bond price is given by a deterministic function $C = C(V, p, t)$;

**A Financial explanation about assumptions**:

Assumption 4) means that we consider the two cases: the first is the case that default boundary is constant and second is the case that default boundary is the discounted value of the maturity default boundary. Assumption 5) means default recovery is the same with paying $R$ at maturity. Here assumption (3.1) is needed *to hedge* the risk of $V$ as in [Li-Ren 2004], but *without it, our arguments still hold, too.*

**The Deriving Pricing PDE Model:**

Here our risk sources V and p are all not traded assets, so we hedge those risks using the traded asset $S$ and the "virtual bond" $\hat{C}(p, t)$ (the corporate zero coupon bond with unexpected default only, considered in the previous section, its default recovery is denoted by $\hat{R}$).

Construct a portfolio:

$$\Pi = C - \Delta_1 S - \Delta_2 \hat{C}$$

so that

$$d\Pi = dC - \Delta_1 dS - \Delta_2 d\hat{C} = r\Pi dt. \tag{3.3}$$

From (3.1) we have

$$\Pi = (1 + \frac{\Delta_1 n}{m})C - \frac{\Delta_1}{m}V - \Delta_2 \hat{C}, \tag{3.4}$$

So

$$d\Pi = (1 + \frac{\Delta_1 n}{m})dC - \frac{\Delta_1}{m}(dV + bdtV) - \Delta_2 d\hat{C} \tag{3.6}$$

(i) The case of no default (unexpected) in time interval [$t$, $t + dt$]:    (with probability $1 - p_t dt$):

By the two dimensional Itô formula, we have

$$d\Pi = (1 + \frac{\Delta_1 n}{m})dC - \frac{\Delta_1}{m}(dV + bdtV) - \Delta_2 d\hat{C}$$

$$= (1 + \frac{\Delta_1 n}{m})\{ \{\frac{\partial C}{\partial t} + \frac{1}{2}[s_p^2 \frac{\partial^2 C}{\partial p^2} + s_V^2 V^2 \frac{\partial^2 C}{\partial V^2}]\}dt + \frac{\partial C}{\partial V}dV + \frac{\partial C}{\partial p}dp \}$$

$$- \frac{\Delta_1}{m}(dV + bdtV) - \Delta_2 \{(\frac{\partial \hat{C}}{\partial t} + \frac{1}{2}s_p^2 \frac{\partial^2 \hat{C}}{\partial p^2})dt + \frac{\partial \hat{C}}{\partial p}dp \}.$$

Here we choose $\Delta_1$ and $\Delta_2$ such that

$$(1+\frac{\Delta_1 n}{m})\frac{\partial C}{\partial V} - \frac{\Delta_1}{m} = 0, \quad (1+\frac{\Delta_1 n}{m})\frac{\partial C}{\partial p} - \Delta_2 \frac{\partial \hat{C}}{\partial p} = 0,$$

that is,

$$\Delta_1 = m\frac{\partial C}{\partial V}\left(1-n\frac{\partial C}{\partial V}\right)^{-1}, \Delta_2 = \frac{\partial C}{\partial p}\left(\frac{\partial \hat{C}}{\partial p}\right)^{-1}\left(1-n\frac{\partial C}{\partial V}\right)^{-1}, (1+\frac{\Delta_1 n}{m}) = \left(1-n\frac{\partial C}{\partial V}\right)^{-1} \quad (3.7)$$

then we have

$$d\Pi = \left(1-n\frac{\partial C}{\partial V}\right)^{-1}\left\{\frac{\partial C}{\partial t}+\frac{1}{2}[s_p^{\ 2}\frac{\partial^2 C}{\partial p^2}+s_V^{\ 2}V^2\frac{\partial^2 C}{\partial V^2}]-bV\frac{\partial C}{\partial V}\right.$$
$$\left.-\frac{\partial C}{\partial p}\left(\frac{\partial \hat{C}}{\partial p}\right)^{-1}[\frac{\partial \hat{C}}{\partial t}+\frac{1}{2}s_p^{\ 2}\frac{\partial^2 \hat{C}}{\partial p^2}]\right\}dt \quad (3.8)$$

(ii) The case of default (unexpected) in time interval $[t, t + dt]$ (with probability $p_t dt$);

In this case Since we have

$$dC = R_d - C, \quad d\hat{C} = \hat{R} - \hat{C},$$

then substitute them and (3.7) into (3.6), then

$$d\Pi = (1+\frac{\Delta_1 n}{m})(R_d - C) - \frac{\Delta_1}{m}(dV + bdtV) - \Delta_2(\hat{R} - \hat{C})$$
$$= \left(1-n\frac{\partial C}{\partial V}\right)^{-1}\left[(R_d - C) - \frac{\partial C}{\partial V}(dV + bdtV) - \frac{\partial C}{\partial p}\left(\frac{\partial \hat{C}}{\partial p}\right)^{-1}(\hat{R}-\hat{C})\right]. \quad (3.9)$$

Thus combine the above two cases (i) and (ii) and then from (3.3)

$$d\Pi = (3.8)\times(1 - p_t dt) + (3.9)\times p_t dt = r \Pi dt.$$

Consider (3.4) and (3.7), then

$$\Pi = \left(1-n\frac{\partial C}{\partial V}\right)^{-1}\left(C - V\frac{\partial C}{\partial V} - \frac{\partial C}{\partial p}\left(\frac{\partial \hat{C}}{\partial p}\right)^{-1}\hat{C}\right),$$

neglect higher order infinitesimal of $dt$ and multiply $1-n\frac{\partial C}{\partial V}$ to both sides, then we have

$$\frac{\partial C}{\partial t}+\frac{1}{2}[s_p^{\ 2}\frac{\partial^2 C}{\partial p^2}+s_V^{\ 2}V^2\frac{\partial^2 C}{\partial V^2}]+(r-b)V\frac{\partial C}{\partial V}-rC+p(R_d - C)$$
$$= \frac{\partial C}{\partial p}\left(\frac{\partial \hat{C}}{\partial p}\right)^{-1}\left(\frac{\partial \hat{C}}{\partial t}+\frac{1}{2}s_p^{\ 2}\frac{\partial^2 \hat{C}}{\partial p^2} - r\hat{C} + p(\hat{R}-\hat{C})\right).$$

Here, if we consider the equation (2.6) for $\hat{C}(p,t)$, then we have

$$\frac{\partial C}{\partial t}+\frac{1}{2}[s_p^{\,2}\frac{\partial^2 C}{\partial p^2}+s_V^{\,2}V^2\frac{\partial^2 C}{\partial V^2}]+(r-b)V\frac{\partial C}{\partial V}+a_p\frac{\partial C}{\partial p}-rC+p(R_d-C)=0. \quad (3.10)$$

*Note* **1**: Here $a_p(p, t)$ is the risk neutral drift of *p* in (2.8) and $(r - b)$ is also the risk neutral drift of *V*. That is, in our equation we can see only risk neutral drifts of *p* and *V* on the risk neutral martingale measure, although we started with the natural measure, this means our (3.10) is an equation of *fair* price. And this means that *in the risk neutral world* the firm value satisfies

$$dV = (r-b)Vdt + s_V V dW_2, \text{ (risk neutral martingale measure)} \quad (3.2)'$$

*Note* **1**: When the assumption (3.1) is not satisfied, using expectation method we can also get (3.10) starting with the stochastic differential equation (2.8) and (3.2)' in risk neutral world.

So considering assumption 5, we have derived the pricing problem: Solve the following problem in the region $\Sigma = \{(V, p, t) : 0 < t < T, p > 0, V > V_b(t)\}$

$$\begin{cases} \frac{\partial C}{\partial t}+\frac{1}{2}[s_p^{\,2}(p,t)\frac{\partial^2 C}{\partial p^2}+s_V^{\,2}V^2\frac{\partial^2 C}{\partial V^2}]+a_p(p,t)\frac{\partial C}{\partial p}+(r-b)V\frac{\partial C}{\partial V} \\ -(r+p)C+e^{-r(T-t)}Rp=0, & (0<t<T,\ V>V_b(t),\ p>0) \\ C(V_b(t),p,t)=e^{-r(T-t)}R, & (0<t<T,\ p>0) \\ C(V,p,T)=1, & (V>V_b(t),\ p>0) \\ C(V\to\infty):\text{bounded}, & (0<t<T,\ p>0) \\ C(p\to\infty):\text{bounded}, & (0<t<T,\ V>V_b(t)) \end{cases} \quad (3.11)$$

This is a **non-homogeneous equation** with **variable coefficients** and **moving barrier**. Generally speaking, such equations have no solution formulae. But under the above assumptions (in particular, 2, 4, 5) and conditions (2.16-1) or (2.16-2) about $s_p$ and $a_p$, we can solve it.

As in the above, if we use numeraire as bank save account, that is, the change of unknown function

$$C = ue^{-r(T-t)}, \quad (3.12)$$

then we get the **non homogeneous** equation with **risk free rate 0**:

$$\begin{cases} \frac{\partial u}{\partial t}+\frac{1}{2}[s_p^{\,2}(p,t)\frac{\partial^2 u}{\partial p^2}+s_V^{\,2}V^2\frac{\partial^2 u}{\partial V^2}]+a_p(p,t)\frac{\partial u}{\partial p}+(r-b)V\frac{\partial u}{\partial V}-p(u-R)=0, \\ & (0<t<T,\ V>V_b(t),\ p>0) \\ u(V_b(t),p,t)=R, & (0<t<T,\ p>0) \\ u(V,p,T)=1, & (V>V_b(t),\ p>0) \\ u(V\to\infty):\text{bounded}, & (0<t<T,\ p>0) \\ u(p\to\infty):\text{bounded}, & (0<t<T,\ V>V_b(t)) \end{cases} \quad (3.13)$$

Use again the change of unknown function,

$$\hat{u} = u - R, \quad (3.14)$$

then we get the **homogeneous** equation with **risk free rate 0**;

$$\begin{cases} \dfrac{\partial \hat{u}}{\partial t} + \dfrac{1}{2}[s_p^{\,2}(p,t)\dfrac{\partial^2 \hat{u}}{\partial p^2} + s_V^{\,2}V^2\dfrac{\partial^2 \hat{u}}{\partial V^2}] + a_p(p,t)\dfrac{\partial \hat{u}}{\partial p} + (r-b)V\dfrac{\partial \hat{u}}{\partial V} - p\hat{u} = 0, \\ \qquad\qquad\qquad\qquad\qquad\qquad (0<t<T,\ V>V_b(t),\ p>0) \\ \hat{u}(V_b(t),p,t) = 0, \qquad\qquad (0<t<T,\ p>0) \\ \hat{u}(V,p,T) = 1-R, \qquad\qquad (V>V_b(t),\ p>0) \\ \hat{u}(V\to\infty): \text{bounded}, \qquad (0<t<T,\ p>0) \\ \hat{u}(p\to\infty): \text{bounded}, \qquad (0<t<T,\ V>V_b(t)) \end{cases} \quad (3.15)$$

Use again the change of unknown function,

$$\hat{u} = W(1-R), \qquad (3.16)$$

then our defaultable bond price is represented by **W** as

$$\begin{aligned} C(V,p,t) &= ue^{-r(T-t)} = (\hat{u}+R)e^{-r(T-t)} = [W(1-R)+R]e^{-r(T-t)} \\ &= W(V,p,t)e^{-r(T-t)} + (1-W(V,p,t))R\cdot e^{-r(T-t)} \end{aligned} \qquad (3.17)$$

That is, the price of our defaultable bond at time *t* can be seen as an expectation of the current value of the bond in the case of no default and the value of the bond in the case of default at time *t*. So *W*(*V*, *p*, *t*) in (3.17) can be regarded as *a probability at time t of no default* in the case of unexpected default intensity is *p* and expected default barrier is *V*<sub>b</sub>. Thus we get the **PDE equation of no default probability** when default intensity is *p* and expected default barrier is *V*<sub>b</sub>.

$$\begin{cases} \dfrac{\partial W}{\partial t} + \dfrac{1}{2}[s_p^{\,2}(p,t)\dfrac{\partial^2 W}{\partial p^2} + s_V^{\,2}V^2\dfrac{\partial^2 W}{\partial V^2}] + a_p(p,t)\dfrac{\partial W}{\partial p} + (r-b)V\dfrac{\partial W}{\partial V} - pW = 0, \\ \qquad\qquad\qquad\qquad\qquad\qquad (0<t<T,\ V>V_b(t),\ p>0) \\ W(V_b(t),p,t) = 0, \qquad\qquad (0<t<T,\ p>0) \\ W(V,p,T) = 1, \qquad\qquad (V>V_b(t),\ p>0) \\ W(V\to\infty): \text{bounded}, \qquad (0<t<T,\ p>0) \\ W(p\to\infty): \text{bounded}, \qquad (0<t<T,\ V>V_b(t)) \end{cases} \quad (3.18)$$

Since the unexpected default related to *p* and the expected default related to *V*<sub>b</sub> is not correlated by assumption 5), we can guess no default probability can be represented as

$$W(V,p,t) = f(V,t)\cdot g(p,t) \qquad (3.19)$$

Substitute (3.19) into (3.18), then we have

$$\left[\dfrac{\partial f}{\partial t} + \dfrac{1}{2}s_V^{\,2}V^2\dfrac{\partial^2 f}{\partial V^2} + (r-b)V\dfrac{\partial f}{\partial V}\right]g + \left[\dfrac{\partial g}{\partial t} + \dfrac{1}{2}s_p^{\,2}(p,t)\dfrac{\partial^2 g}{\partial p^2} + a_p(p,t)\dfrac{\partial g}{\partial p} - pg\right]f = 0\cdot$$

So if we solve the following two problems, then we get the solution of (3.18).

$$\begin{cases} \dfrac{\partial f}{\partial t} + \dfrac{1}{2}s_V^{\,2}V^2\dfrac{\partial^2 f}{\partial V^2} + (r-b)V\dfrac{\partial f}{\partial V} = 0, \quad (0<t<T,\ V>V_b(t)) \\ f(V_b(t),t) = 0, \qquad\qquad (0<t<T) \\ f(V,T) = 1, \qquad\qquad (V>V_b(t)) \\ f(V\to\infty): \text{bounded}, \qquad (0<t<T) \end{cases} \quad (3.20)$$

$$\begin{cases} \dfrac{\partial g}{\partial t} + \dfrac{1}{2} s_p^{\,2}(p,t) \dfrac{\partial^2 g}{\partial p^2} + a_p(p,t) \dfrac{\partial g}{\partial p} - pg = 0, & (0 < t < T, p > 0) \\ g(p,T) = 1, & (p > 0) \end{cases} \quad (3.21)$$

**Remark**: Thus, we can regard that $g(p, t)$ is a no default probability with unexpected default intensity $p_t$ and $f(V, t)$ is a no default probability with expected default barrier $V_b$.

Since (3.21) is just the same with (2.14), we can easily solve it under the conditions (2-16) and (2-18) about $s_p$ and $a_p$, the solution of (3.21) is:

$$g(p,t) = e^{A(t,T) - B(t,T)p}. \quad (3.22)$$

where $A(t, T)$ is given by (2.22) and $B(t, T)$ is given by

$$B(t,T) = \begin{cases} \dfrac{1 - \exp(-c(T-t))}{c}, & dp_t = (b(t) - c \cdot p)dt + \sqrt{d(t)} \cdot dW, c \neq 0, \\ T - t, & dp_t = b(t)dt + \sqrt{d(t)} \cdot dW, \\ \sqrt{2/C}\,\text{th}[\sqrt{C/2}(T-t)], & dp_t = b(t)dt + \sqrt{d(t) + C \cdot p} \cdot dW, C > 0. \end{cases} \quad (3.23)$$

(3.20) can be thought as a pricing equation of a "**bond with barrier** with *risk free rate* 0 and *dividend rate* $(b - r)$, $a = 0$ or $r$". We can easily get the solution using the same method used in put-call parity for down and out barrier options (see ch. 8, theorem **1** of [Jiang LS 03] in the case of $V_b(t) \equiv$ const and the formula (32) of [OHC 2004] in the case of $V_b(t) = e^{-r(T-t)}$).

$$f(V,t) = \begin{cases} N(d_1) - \left(\dfrac{V}{V_B}\right)^{1 - \frac{2(r-b)}{s_V^2}} N(d_2), & V_b(t) \equiv V_B : \text{const}, \\ N(d_1') - \left(\dfrac{V}{V_B}\right)^{1 + \frac{2b}{s_V^2}} e^{(1 + \frac{2b}{s_V^2})r(T-t)} N(d_2'), & V_b(t) \equiv V_B e^{-r(T-t)}. \end{cases} \quad (3.24)$$

where

$$d_1 = \dfrac{\ln \dfrac{V}{V_B} + (r - b - \dfrac{s_V^2}{2})(T-t)}{s_V \sqrt{T-t}}, \quad d_2 = \dfrac{\ln \dfrac{V_B}{V} + (r - b - \dfrac{s_V^2}{2})(T-t)}{s_V \sqrt{T-t}}$$

$$d_1' = \dfrac{\ln \dfrac{V}{V_B} + (r - b - \dfrac{s_V^2}{2})(T-t)}{s_V \sqrt{T-t}}, \quad d_2' = \dfrac{\ln \dfrac{V_B}{V} + (-r - b - \dfrac{s_V^2}{2})(T-t)}{s_V \sqrt{T-t}}$$

Thus we have the **no default probability formula** and the **pricing formula** of *defaultable zero coupon corporate bond* with **stochastic unexpected default intensity** and expected **default barrier**:

$$W(V,p,t) = \begin{cases} e^{A(t,T)-B(t,T)p}\left[N(d_1) - \left(\dfrac{V}{V_B}\right)^{1-\frac{2(r-b)}{s_V^2}} N(d_2)\right], & V_b(t) \equiv V_B : \text{const}, \\ e^{A(t,T)-B(t,T)p}\left[N(d_1') - \left(\dfrac{V}{V_B}\right)^{1+\frac{2b}{s_V^2}} e^{(1+\frac{2b}{s_V^2})r(T-t)} N(d_2')\right], & V_b(t) \equiv V_B e^{-r(T-t)}. \end{cases}$$

(3.25)

$$C(V,p,t) = R \cdot e^{-r(T-t)} + W(V,p,t)(1-R)e^{-r(T-t)}. \tag{3.26}$$

**Remark**: The formulae (3.25) and (3.26) are also true in the case when the default intensity $p_t$ satisfies the models (2.16) but do not satisfy one of the conditions of (2.18), although in this case $B(t, T)$ cannot have an explicit formula.

The *financial meaning* of (3.26) is clear: the *first term* of (3.26) is the current price of *the part to be given to bond holder **regardless of default occurs or not***, and the *second term* is the allowance dependent on default probability at time **t**. If at some moment **t**, the default is certain ($W(V, p, t) = 0$), then the price of the bond at **t** is exactly the current price of default recovery. And as shown in formula (3.17) or (3.26), if $R = 0$, that is, if there is nothing to recover in case of default, then the bond price is the product of zero coupon bond and survival probability; and if $R = 1$, that is, if there is no any loss in case of default, then the bond price is the same with zero coupon bond.

Using the same method, we can derive a ***price formula of a credit default swap***. Consider a credit default swap which pays $(1-R)\cdot e^{-r(T-t)}$ in case of unexpected default and the expected default before time *T* to the corporate bond holder and it require the protection buyer must pay the price of the credit default swap at the first date of that insurance contract. Let denote our swap by $S(V, p, t)$, then it satisfies the following problem in the region

$$\Sigma = \{(V,p,t) : 0 < t < T, p > 0, V > V_b(t)\}.$$

$$\begin{cases} \dfrac{\partial S}{\partial t} + \dfrac{1}{2}\left[s_p^2(p,t)\dfrac{\partial^2 S}{\partial p^2} + s_V^2 V^2 \dfrac{\partial^2 S}{\partial V^2}\right] + a_p(p,t)\dfrac{\partial S}{\partial p} + (r-b)V\dfrac{\partial S}{\partial V} \\ \quad -(r+p)S + e^{-r(T-t)}(1-R)p = 0, & (0 < t < T, \ V > V_b(t), \ p > 0) \\ S(V_b(t), p, t) = e^{-r(T-t)}(1-R), & (0 < t < T, \ p > 0) \\ S(V, p, T) = 0, & (V > V_b(t), \ p > 0) \\ S(V \to \infty): \text{bounded}, & (0 < t < T, \ p > 0) \\ S(p \to \infty): \text{bounded}, & (0 < t < T, \ V > V_b(t)) \end{cases} \tag{3.27}$$

Using the same method as the above, we can get

$$S(V,p,t) = (1-W)(1-R)e^{-r(T-t)}, \tag{3.28}$$

Here *W* is the no default probability given by (3.25).

The *financial meaning* of (3.28) is clear: since the swap contract gives nothing in case of no default and gives compensation $(1-R)\cdot e^{-r(T-t)}$ in case of default, then the price of the swap contract is the product of the compensation and default probability.

***Remark 1.*** The assumption 3 in section 2 and assumption 4, 5 in section 3 (about default barrier and recovery) made an important role in deriving analytical pricing formula. If the default recovery value was not a discounted value but a constant, then it is rather difficult to change non homogeneous equation (2.9) or (3.11) into homogeneous equation (2.12) or (3.15). And if expected default recovery and unexpected default recovery were not same, then our logic would not work and it would need another method (for example, [Realdon 2004]).

***Remark 2.*** In this section, the assumption (3.1) that the firm value is sum of some shares of stocks and defaultable bonds made an important role in hedging the risk of firm value ***V***. Here, we hedged the risk of firm value ***V*** using (3.1) and the unexpected default risk ***p*** using the "virtual bond" with only unexpected default, so our equation (3.18) has no any drifts on natural measure and it gives fair price to our bonds.

***Remark 3.*** In the case when the drift and squared volatility of default intensity $p_t$ are linear on ***p*** but do not satisfy one of the conditions (2.18), we have ***semi-analytical formula*** of bond price, because $B(t, T)$ in pricing formulae cannot have an explicit formula.

## $ 4. The Pricing Corporate Zero Coupon Bond with Unexpected Default only -The Case with Stochastic Short Rate.

In order to clarify the effect of stochastic unexpected default intensity, we consider the simplest case that the price of corporate zero coupon bond is not effected by the firm value but only effected by unexpected default.

Main Assumptions are as follows:

**Assumption** 1: Risk free rate follows *Vasicek model*:

$$dr_t = \theta(\mu_r - r_t)dt + s_r dW_1(t) \text{ (risk neutral martingale measure)} \quad (4.1)$$

**Assumption** 2: Unexpected default probability in $[t, t + dt]$ is $p_t\,dt$, the default intensity $p_t$ satisfies

$$dp = a_p(r,p,t)dt + s_p(r,p,t)dW_2, \text{(risk neutral martingale measure)} \quad (4.2)$$

$$a_p(r,p,t) = \alpha(t) + \beta(t)r + \chi(t)p,$$
$$s_p^2(r,p,t) = \delta(t) + \varepsilon(t)r + \phi(t)p. \quad (4.3)$$

**Assumption** 3: Unexpected default recovery is given as the following two forms
  (i) Face value recovery: $R_d = R \cdot Z$; $R: 0 \leq R \leq 1$:const
      here ***Z*** is the price at the default time of default free zero coupon bond.
  (ii): Market price recovery: $R_d = R \times$(bond price at the default time), $0 \leq R \leq 1$:const

**Assumption** 4: Covariances:

$$Cov(dW_1, dW_2) = \rho \quad (4.4)$$

Furthermore, after deriving the pricing equation (4.13), we assume that

$$\rho = 0, \quad (4.5)$$

that is, <u>unexpected default is not correlated with the short rate</u> when deriving the pricing formula. (This is not seemed to be a *likable assumption* but we need it to get *analytical pricing formula*);

**Assumption** 5: The price of our corporate defaultable zero coupon bond with unexpected default only is given by a deterministic function $\hat{C} = \hat{C}(r, p, t)$

**A Financial explanation about assumptions**:

We suppose that such a model is not only useful when we do not know the circumstance of the firm value but also used as a "***virtual bond***" ***to hedge the risk*** of unexpected default even when we know very well about the circumstance of the firm value, as shown in next section.

**The Deriving of PDE model.**

Under the condition (4.1), the price $Z(r, t; T)$ of default free zero coupon bond is given as follows:

$$\begin{cases} \frac{\partial Z}{\partial t} + \frac{1}{2} s_r^2 \frac{\partial^2 Z}{\partial r^2} + \theta(\mu_r - r_t)\frac{\partial Z}{\partial r} - rZ = 0, \\ Z(r,T) = 1, \end{cases} \quad (4.6)$$

and

$$Z(r,t;T) = e^{\bar{A}(t,T) - \bar{B}(t,T)r}$$
$$\bar{B}(t,T) = \frac{1 - e^{-\theta(T-t)}}{\theta}. \quad (4.7)$$

Here, we have no any underlying assets with which to hedge the risk of $p_t$. Beside of default free zero coupon bond, the only traded asset is our bond as a derivative of our independent variable $p_t$ and so the only way to construct a hedged portfolio is by hedging one bond with a bond of a different maturity [Wilmott 98].

Let denote the price of a bond with maturity $T_i$ and default recovery $R_i$ by $C_i(r, p, t; T_i)$, i = 1, 2. Construct a portfolio:

$$\Pi = C_1 - \Delta_1 Z - \Delta_2 C_2$$

such that it is risk-free, that is,

$$d\Pi = dC_1 - \Delta_1 dZ - \Delta_2 dC_2 = r\Pi dt \quad (4.8)$$

(i) The case of no default (with probability $1 - p_t dt$)

$$d\Pi = \left( \frac{\partial C_1}{\partial t} + \frac{1}{2}[s_r^2 \frac{\partial^2 C_1}{\partial r^2} + 2\rho s_r s_p \frac{\partial^2 C_1}{\partial r \partial p} + s_p^2 \frac{\partial^2 C_1}{\partial p^2}] \right) dt + \frac{\partial C_1}{\partial r} dr + \frac{\partial C_1}{\partial p} dp$$

$$- \Delta_1 \left[ \left( \frac{\partial Z}{\partial t} + \frac{1}{2} s_r^2 \frac{\partial^2 Z}{\partial r^2} \right) dt + \frac{\partial Z}{\partial r} dr \right]$$

$$- \Delta_2 \left[ \left( \frac{\partial C_2}{\partial t} + \frac{1}{2}[s_r^2 \frac{\partial^2 C_2}{\partial r^2} + 2\rho s_r s_p \frac{\partial^2 C_2}{\partial r \partial p} + s_p^2 \frac{\partial^2 C_2}{\partial p^2}] \right) dt + \frac{\partial C_2}{\partial r} dr + \frac{\partial C_2}{\partial p} dp \right]$$

$$= \left(\frac{\partial C_1}{\partial t} + \frac{1}{2}[s_r^2 \frac{\partial^2 C_1}{\partial r^2} + 2\rho s_r s_p \frac{\partial^2 C_1}{\partial r \partial p} + s_p^2 \frac{\partial^2 C_1}{\partial p^2}]\right)dt - \Delta_1\left(\frac{\partial Z}{\partial t} + \frac{1}{2}s_r^2 \frac{\partial^2 Z}{\partial r^2}\right)dt$$

$$-\Delta_2\left(\frac{\partial C_2}{\partial t} + \frac{1}{2}[s_r^2 \frac{\partial^2 C_2}{\partial r^2} + 2\rho s_r s_p \frac{\partial^2 C_2}{\partial r \partial p} + s_p^2 \frac{\partial^2 C_2}{\partial p^2}]\right)dt$$

$$+ (\frac{\partial C_1}{\partial r} - \Delta_1 \frac{\partial Z}{\partial r} - \Delta_2 \frac{\partial C_2}{\partial r})dr + (\frac{\partial C_1}{\partial p} - \Delta_2 \frac{\partial C_2}{\partial p})dp.$$

So let

$$\frac{\partial C_1}{\partial r} - \Delta_1 \frac{\partial Z}{\partial r} - \Delta_2 \frac{\partial C_2}{\partial r} = 0, \quad \frac{\partial C_1}{\partial p} - \Delta_2 \frac{\partial C_2}{\partial p} = 0,$$

that is,

$$\Delta_1 = \left[\frac{\partial C_1}{\partial r} - \frac{\partial C_1}{\partial p}\left(\frac{\partial C_2}{\partial p}\right)^{-1}\frac{\partial C_2}{\partial r}\right]\left(\frac{\partial Z}{\partial r}\right)^{-1}, \quad \Delta_2 = \frac{\partial C_1}{\partial p}\left(\frac{\partial C_2}{\partial p}\right)^{-1}, \tag{4.9}$$

then

$$d\Pi = \left\{\frac{\partial C_1}{\partial t} + \frac{1}{2}[s_r^2 \frac{\partial^2 C_1}{\partial r^2} + 2\rho s_r s_p \frac{\partial^2 C_1}{\partial r \partial p} + s_p^2 \frac{\partial^2 C_1}{\partial p^2}] - \Delta_1\left(\frac{\partial Z}{\partial t} + \frac{1}{2}s_r^2 \frac{\partial^2 Z}{\partial r^2}\right)\right.$$
$$\left. - \Delta_2\left(\frac{\partial C_2}{\partial t} + \frac{1}{2}[s_r^2 \frac{\partial^2 C_2}{\partial r^2} + 2\rho s_r s_p \frac{\partial^2 C_2}{\partial r \partial p} + s_p^2 \frac{\partial^2 C_2}{\partial p^2}]\right)\right\}dt \tag{4.10}$$

(ii) The case of default (with probability $p_t dt$)

$$d\Pi = (R_1 - C_1) - \Delta_1 dZ - \Delta_2(R_2 - C_2), \tag{4.11}$$

If we combine (i) and (ii), $d\Pi = (4.10)\times(1-p_t dt)+(4.11)\times p_t dt = r\Pi dt$, and neglecting the higher order infinitesimal of $dt$, then we have

$$\frac{\partial C_1}{\partial t} + \frac{1}{2}[s_r^2 \frac{\partial^2 C_1}{\partial r^2} + 2\rho s_r s_p \frac{\partial^2 C_1}{\partial r \partial p} + s_p^2 \frac{\partial^2 C_1}{\partial p^2}] - rC_1 + p(R_1 - C_1)$$
$$-\Delta_2\left\{\frac{\partial C_2}{\partial t} + \frac{1}{2}[s_r^2 \frac{\partial^2 C_2}{\partial r^2} + 2\rho s_r s_p \frac{\partial^2 C_2}{\partial r \partial p} + s_p^2 \frac{\partial^2 C_2}{\partial p^2}] - rC_2 -_2 p(R_2 - C_2)\right\}$$
$$-\Delta_1\left(\frac{\partial Z}{\partial t} + \frac{1}{2}s_r^2 \frac{\partial^2 Z}{\partial r^2} - rZ\right) = 0.$$

Here consider (4.6) and (4.9), then

$$\left\{\frac{\partial C_1}{\partial t} + \frac{1}{2}[s_r^2 \frac{\partial^2 C_1}{\partial r^2} + 2\rho s_r s_p \frac{\partial^2 C_1}{\partial r \partial p} + s_p^2 \frac{\partial^2 C_1}{\partial p^2}] + \theta(\mu_r - r)\frac{\partial C_1}{\partial r} - rC_1 + p(R_1 - C_1)\right\}\left(\frac{\partial C_1}{\partial p}\right)^{-1} =$$
$$= \left\{\frac{\partial C_2}{\partial t} + \frac{1}{2}[s_r^2 \frac{\partial^2 C_2}{\partial r^2} + 2\rho s_r s_p \frac{\partial^2 C_2}{\partial r \partial p} + s_p^2 \frac{\partial^2 C_2}{\partial p^2}] + \theta(\mu_r - r)\frac{\partial C_2}{\partial r} - rC_2 + p(R_2 - C_2)\right\}\left(\frac{\partial C_2}{\partial p}\right)^{-1}.$$

The left hand of this equation is a function of $T_1$ but not $T_2$, and the right hand is a a function of $T_2$ but not $T_1$, so the both sides must be a function $-a(r, p, t)$ independent of $T_1, T_2$. Thus we have

$$\frac{\dfrac{\partial C}{\partial t} + \dfrac{1}{2}[s_r^2 \dfrac{\partial^2 C}{\partial r^2} + 2\rho s_r s_p \dfrac{\partial^2 C}{\partial r \partial p} + s_p^2 \dfrac{\partial^2 C}{\partial p^2}] + \theta(\mu_r - r)\dfrac{\partial C}{\partial r} - rC + p(R - C)}{\dfrac{\partial C}{\partial p}} = -u(r, p, t).$$

Let

$$\lambda(r,p,t) = \frac{a_p(r,p,t) - u(r,p,t)}{s_p(r,p,t)},$$

that is,

$$u(r,p,t) = a_p(r,p,t) - s_p(r,p,t)\lambda(r,p,t), \tag{4.12}$$

then $\lambda(r, p, t)$ is called a **market price of risk** $p_t$ and $u(r, p, t)$ is called a **risk neutral drift** of $p_t$, that is, $u(r, p, t)$ is the coefficient of **dt** in stochastic differential equation for *p* in the risk neutral martingale measure)[Wilmott 1998], so from the assumption (4.3),

$$u(r,p,t) = a_p(r,p,t),$$

Thus we have the PDE model of the price $\hat{C} = \hat{C}(r,p,t)$ of our corporate defaultable zero coupon bond with unexpected default only (default intensity $p_t$ and recovery $R_d$):

$$\begin{cases} \frac{\partial \hat{C}}{\partial t} + \frac{1}{2}[s_r^2 \frac{\partial^2 \hat{C}}{\partial r^2} + 2\rho s_r s_p \frac{\partial^2 \hat{C}}{\partial r \partial p} + s_p^2 \frac{\partial^2 \hat{C}}{\partial p^2}] + \theta(\mu_r - r)\frac{\partial \hat{C}}{\partial r} + a_p \frac{\partial \hat{C}}{\partial p} \\ \quad - (r+p)\hat{C} + pR_d = 0, \quad (0 < t < T, r > 0, p > 0,) \\ \hat{C}(r,p,T) = 1, \quad (r > 0, p > 0) \end{cases} \tag{4.13}$$

**Solving:**

**4.1 *The Case of Face Value Recovery*** ($R_d = R \cdot Z$; $R$: $0 \leq R \leq 1$:const);

Then (4.13) is changed to

$$\begin{cases} \frac{\partial \hat{C}}{\partial t} + \frac{1}{2}[s_r^2 \frac{\partial^2 \hat{C}}{\partial r^2} + s_p^2 \frac{\partial^2 \hat{C}}{\partial p^2}] + \theta(\mu_r - r)\frac{\partial \hat{C}}{\partial r} + a_p \frac{\partial \hat{C}}{\partial p} - (r+p)\hat{C} \\ \quad + pRZ(r,t) = 0, \quad (0 < t < T, r > 0, p > 0,) \\ \hat{C}(r,p,T) = 1, \quad (r > 0, p > 0) \end{cases} \tag{4.14}$$

In order to get solution formula, we assume that the ***drift*** and ***volatility*** of *p* is not related to short rate *r* , that is,

$$\beta(t) = \varepsilon(t) = 0 \text{ in (4.3)}, \tag{4.3'}$$
$$\rho = 0$$

Use numeraire as zero coupon bond price **Z**, that is, the change of unknown function and variable

$$\hat{C}(r,p,t) = Z(r,t) \cdot u(p,t), \tag{4.15}$$

consider (4.7) and divide both sides by **Z**, then we get the ***non homogeneous*** equation with ***linear non homogeneous term Rp*** and ***risk free rate* 0**:

$$\begin{cases} u_t + \frac{1}{2} s_p^2(p,t) u_{pp} + a_p(p,t) u_p - p(u-R) = 0, & (p > 0, \ 0 < t < T) \\ u(p,T) = 1, & (p > 0) \end{cases} \quad (4.16)$$

Use again the change of unknown function,

$$\hat{u} = u - R, \quad (4.17)$$

then we get the *homogeneous* equation with *risk free rate* **0**;

$$\begin{cases} \hat{u}_t + \frac{1}{2} s_p^2(p,t) \hat{u}_{pp} + a_p(p,t) \hat{u}_p - p\hat{u} = 0, & (p > 0, \ 0 < t < T) \\ \hat{u}(p,T) = 1 - R, & (p > 0) \end{cases} \quad (4.18)$$

Use again the change of unknown function,

$$\hat{u} = W(1-R), \quad (4.19)$$

then our defaultable bond price is represented by **W** as

$$\begin{aligned}\hat{C}(r,p,t) &= uZ = (\hat{u}+R)Z = Z[W(1-R)+R] \\ &= Z(r,t)\{W(p,t) + [1-W(p,t)]R\}\end{aligned} \quad (4.20)$$

That is, the price of our defaultable bond at time *t* can be seen as an expectation of the current value of the bond in the case of no default and the value of the bond in the case of default at time *t*. So $W(p, t)$ in (4.20) can be regarded as *a probability at time t of no default* in the case of unexpected default intensity is $p_t$. Thus we get the **PDE equation of no default probability** when default intensity is $p_t$

$$\begin{cases} W_t + \frac{1}{2} s_p^2(p,t) W_{pp} + a_p(p,t) W_p - pW = 0, & (p > 0, \ 0 < t < T) \\ W(p,T) = 1, & (p > 0) \end{cases} \quad (4.21)$$

Then under the assumption (4.3) and (4.3′) we have the solution of (4.21) in the form of

$$W(p,t) = e^{A(t,T) - B(t,T) p}. \quad (4.22)$$

Here if

$$\begin{aligned} a_p(p,t) &= \alpha(t) + \chi(t) p, \\ s_p^2(r,p,t) &= \delta(t) + \phi(t) p, \end{aligned} \quad (4.23)$$

then $B(t, T)$ and $A(t, T)$ are the solution of the following ordinary differential equations:

$$B' + \chi(t) B - \frac{1}{2} \phi(t) B^2 + 1 = 0. \quad (4.24)$$

$$A' + \frac{1}{2} \delta(t) B^2 - \alpha(t) B = 0 \quad (4.25)$$

From $W(p, T) = 1$, we know that

$$A(T, T) = B(T, T) = 0$$

As shown in section 2, (4.24) has the following solutions under the some restrictions on the coefficients for (4.23):

$$B(t,T) = \begin{cases} T-t, & dp_t = \alpha(t)dt + \sqrt{\delta(t)} \cdot dW, \\ \dfrac{1-\exp(-c(T-t))}{c}, & dp_t = (\alpha(t) - c \cdot p)dt + \sqrt{\delta(t)} \cdot dW, c \neq 0 : \text{const}, \\ \sqrt{2/C}\,\text{th}[\sqrt{C/2}(T-t)], & dp_t = \alpha(t)dt + \sqrt{\delta(t) + C \cdot p} \cdot dW, C > 0 : \text{const}. \end{cases} \quad (4.26)$$

Here

$$\text{th}(x) = \frac{e^x - e^{-x}}{e^x + e^{-x}}.$$

(These cases include *Vasicek model* ($\alpha$, $\chi$, $\delta$: const, $\phi = 0$), *Ho-Lee model* ($\chi = 0$, $\delta$: const, $\phi = 0$) and *Hull-White model* ($\chi$, $\delta$: const, $\phi = 0$) ). Once $B(t)$ is known, then $A(t)$ is given by

$$A(t,T) = -\int_t^T [\alpha(s)B(s) - \frac{1}{2}\delta(s)B^2(s)]ds \quad (4.27)$$

Thus, In the case when the default intensity $p_t$ satisfies the condition of (4.26), the **price of corporate bond with unexpected default** intensity $p_t$ is given by

$$\hat{C}(r,p,t) = Z(r,t)\left[ e^{A(t,T)-B(t,T)p} + (1-e^{A(t,T)-B(t,T)p})R \right], \quad (4.28)$$

$$= Z(r,t)\left[ R + (1-R)e^{A(t,T)-B(t,T)p} \right], \quad (4.29)$$

where $B(t, T)$ and $A(t, T)$ are respectively given by (4.26) and (4.27).

***Remark* 1**: The formulae (4.28) and (4.29) are also true in the case when the default intensity $p_t$ satisfies the models (4.23) but not the conditions in (4.26), although in this case $B(t, T)$ cannot have an explicit formula.

***Remark* 2**: The *financial meaning* of (4.28) is clear as mentioned in the above. The *financial meaning* of (4.29) is as follows: the *first term* of (4.29) is the current price of *the part to be given to bond holder* **regardless of default occurs or not**, and the *second term* is the allowance dependent on default probability at time *t*. If at some moment *t*, the *default is certain* ($W(p, t) = 0$), then the price of the bond at *t* is exactly the current price of default recovery. If default *recovery is zero*, that is, $R = 0$, then the ratio of the defaultable bond price and default free zero coupon bond price is the very no default probability. If default *recovery is full*, that is, $R = 1$, then default event does not effect to the bond price and defaultable bond price is the same with default free zero coupon bond price.

**4.2 *The Case of Market price Recovery*** ($R_d = R \cdot C$; $R$: $0 \leq R \leq 1$:const);

Then (4.13) is changed to

$$\begin{cases} \dfrac{\partial \hat{C}}{\partial t} + \dfrac{1}{2}[s_r^2 \dfrac{\partial^2 \hat{C}}{\partial r^2} + s_p^2 \dfrac{\partial^2 \hat{C}}{\partial p^2}] + \theta(\mu_r - r)\dfrac{\partial \hat{C}}{\partial r} + a_p \dfrac{\partial \hat{C}}{\partial p} - (r + p(1-R))\hat{C} = 0, \\ \hat{C}(r,p,T) = 1, \quad (r > 0, p > 0, 0 < t < T) \end{cases} \quad (4.30)$$

Here, in order to get solution formula, we assumed that

$$\rho = 0.$$

Then we can find the solution of (4.30) in the form of

$$\hat{C}(r, p, t) = e^{A(t,T) - B(t,T)r - C(t,T)p}. \tag{4.31}$$

Generally, in (4.31) it is impossible to find $C$, $B$, $A$ explicitly, but *in 3 important cases, we can find them* and $C$, $B$, $A$ are respectively given as follows: (see appendix 2).

$$C(t,T) = \begin{cases} \dfrac{R - 1 + (1 - R)\exp(c(T-t))}{c}, \text{ when} \\ \qquad dp = [\alpha(t) + \beta(t)r + c \cdot p]dt + \sqrt{\delta(t) + \varepsilon(t)r} \cdot dW, c \neq 0, \\ (1-R)(T-t), \text{ when } dp = [\alpha(t) + \beta(t)r]dt + \sqrt{\delta(t) + \varepsilon(t)r} \cdot dW \\ \sqrt{\dfrac{2(1-R)}{c}} \operatorname{th}\left[\sqrt{\dfrac{(1-R)c}{2}}(T-t)\right], \text{ when} \\ \qquad dp = [\alpha(t) + \beta(t)r]dt + \sqrt{\delta(t) + \varepsilon(t)r + c \cdot p} \cdot dW, c > 0. \end{cases} \tag{4.32}$$

$$B(t,T) = -e^{-\theta(T-t)}\left[\int_t^T (\frac{1}{2}\varepsilon(u)C^2(u) - \beta(u)C(u) - 1)e^{\theta(T-u)}du\right], \tag{4.33}$$

$$A(t,T) = \int_t^T [\theta\mu_r B(u) + \alpha(u)C(u) - \frac{1}{2}s_r^2 \cdot B^2(u) - \frac{1}{2}\delta(u)C^2(u)]du, \tag{4.34}$$

Then our defaultable bond price is represented as (4.31).

Let denote the no default probability by **W**. In the case of market price recovery, default recovery is $R\hat{C}(r, p, t)$, and thus

$$\begin{aligned}\hat{C}(r, p, t) &= WZ(r,t) + [1-W]R\hat{C}(r, p, t) \\ &= R\hat{C}(r, p, t) + W[Z(r,t) - R\hat{C}(r, p, t)]\end{aligned} \tag{4.35}$$

Thus no default probability at *t* is given by

$$W = W(r, p, t) = \begin{cases} \dfrac{\hat{C}(r, p, t) - R\hat{C}(r, p, t)}{Z(r,t) - R\hat{C}(r, p, t)}, & 0 \leq R < 1, \\ 1, & R = 1, \end{cases} \tag{4.36}$$

***Remark* 3**: The formulae (4.31), (4.35) are also true in the case when the default intensity $p_t$ satisfies the models (4.3) but does not satisfy one of the conditions in (4.32), although in this case $C(t, T)$ cannot have an explicit formula like (4.32).

***Remark* 4**: The ***financial meaning*** of (4.35) is as follows: the *second term* of (4.35) is the current price of *the part to be given to bond holder **regardless of default occurs or not***, and the *second term* is the allowance dependent on default probability at time ***t***. If at some moment ***t***, the <u>default is certain</u> ($W(p, t) = 0$), then the price of the bond at ***t*** is exactly the current price of default recovery. If default <u>recovery is zero</u>, that is, $R = 0$, then the ratio of the defaultable bond price and default free zero coupon bond price is the very no default probability. If default <u>recovery is full</u>, that is, $R = 1$, then $C(t, T) = 0$ in (4.32) and default event does not effect to the bond price and defaultable bond price is the same with default free zero coupon bond price.

*Note*: [Wilmott1998] noted that (4.14) has the solution of the form of (4.15), (4.31) when default recovery = 0.

## $ 5. The Pricing Corporate Zero Coupon Bond with Both Expected and Unexpected Defaults - The Case with Stochastic Short Rate.

In this section, we studied the pricing formula of the Corporate Zero Coupon Bond with both of expected default and unexpected default. Here we used the "Corporate Zero Coupon Bond with Unexpected Default only" (considered in previous section) as a "*virtual bond*" *to hedge the risk* of unexpected default in deriving the pricing PDE.

Main Assumptions are as follows:

**Assumption 1**: Risk free rate follows *Vasicek model*:

$$dr_t = \theta(\mu_r - r_t)dt + s_r dW_1(t). \tag{5.1}$$

**Assumption 2**: The firm value *V* consists of *m* shares of **traded** assets *S* and *n* sheets of *zero coupon bonds*:

$$V_t = mS_t + nC_t \tag{5.2}$$

and follows the geometric Brownian motion (volatility $s_V$ are assumed to be constant):

$$dV(t) = rV(t)dt + s_V V(t)dW_2(t) \quad (\text{risk neutral measure}) \tag{5.3}$$

**Assumption 3:** Expected default occurs when

$$V \leq V_b(r,t); \quad V_b(r,t) = V_B Z(r,t), \quad V_B : \text{const},$$

where $Z(r, t)$ is the price of default free zero coupon bond.

**Assumption 4:** Unexpected default probability : in $[t, t + dt]$ is $p_t dt$: the default intensity $p_t$ satisfies

$$dp = a_p(p,t)dt + s_p(p,t)dW_3, (\text{risk neutral measure}) \tag{5.4}$$

$$\begin{aligned} a_p(p,t) &= b(t) + c(t)p, \\ s_p^2(p,t) &= d(t) + e(t)p. \end{aligned} \tag{5.5}$$

**Assumption 5:** Expected and unexpected default recovery is exogenous face value recovery:

$$R_d = R \cdot Z; \, R: 0 \leq R \leq 1: \text{const},$$

**Assumption 6:**

$$dW_i \cdot dW_j = \rho_{ij} \, dt, \quad i = 1, 2, 3 \tag{5.6}$$

Furthermore, we assume that $\rho_{23} = 0$, that is, <u>unexpected default is not correlated with firm value</u>(this is a very likable assumption) and, after deriving the pricing quation (5.18) and no default probability equation (5.26), we assume that $\rho_{13} = 0$, that is, <u>unexpected default is not correlated with the short rate</u> when deriving the pricing formula. (This is not seemed to be a *likable assumption* but we need it to get *analytical pricing formula*);

**Assumption 7:** Corporate defaultable zero coupon bond price is given by a deterministic function

$$C = C(r, V, p, t)$$

**PDE Model**

Here our risk sources *r*, *V* and *p* are all not traded assets, so we hedge those risks using the traded asset *S*, default free zero coupon bond *Z*(*r*, *t*) and the "virtual bond" $\hat{C}(r, p, t)$ (the corporate zero coupon bond with unexpected default only, considered in the previous section, its default recovery is denoted by $\hat{R}$).

Construct a portfolio:

$$\Pi = C - \Delta_1 S - \Delta_2 \hat{C} - \Delta_3 Z$$

so that

$$d\Pi = dC - \Delta_1 dS - \Delta_2 d\hat{C} - \Delta_3 dZ = r\Pi dt. \tag{5.7}$$

From (5.2) we have

$$\Pi = (1 + \frac{\Delta_1 n}{m})C - \frac{\Delta_1}{m}V - \Delta_2 \hat{C} - \Delta_3 Z, \tag{5.8}$$

So

$$d\Pi = (1 + \frac{\Delta_1 n}{m})dC - \frac{\Delta_1}{m}dV - \Delta_2 d\hat{C} - \Delta_3 dZ \tag{5.9}$$

(i) The case of no default (unexpected) in time interval [*t*, *t* + *dt*]: (with probability $1 - p_t dt$): By the two dimensional Itô formula, we have

$$\begin{aligned}
d\Pi &= (1+\frac{\Delta_1 n}{m})dC - \frac{\Delta_1}{m}dV - \Delta_2 d\hat{C} \\
&= (1+\frac{\Delta_1 n}{m})\{ \{\frac{\partial C}{\partial t} + \frac{1}{2}[s_r^2 \frac{\partial^2 C}{\partial r^2} + 2\rho_{12} s_r s_V V \frac{\partial^2 C}{\partial r \partial V} + s_V^2 V^2 \frac{\partial^2 C}{\partial V^2} \\
&\quad + 2\rho_{13} s_r s_p \frac{\partial^2 C}{\partial r \partial p} + s_p^2 \frac{\partial^2 C}{\partial p^2}]\}dt + \frac{\partial C}{\partial r}dr + \frac{\partial C}{\partial V}dV + \frac{\partial C}{\partial p}dp \} \\
&\quad - \frac{\Delta_1}{m}dV - \Delta_2\{(\frac{\partial \hat{C}}{\partial t} + \frac{1}{2}[s_r^2 \frac{\partial^2 \hat{C}}{\partial r^2} + 2\rho_{13} s_r s_p \frac{\partial^2 \hat{C}}{\partial r \partial p} \\
&\quad + s_p^2 \frac{\partial^2 \hat{C}}{\partial p^2}])dt + \frac{\partial \hat{C}}{\partial r}dr + \frac{\partial \hat{C}}{\partial p}dp\} - \Delta_3\{(\frac{\partial Z}{\partial t} + \frac{1}{2}s_r^2 \frac{\partial^2 Z}{\partial r^2})dt + \frac{\partial Z}{\partial r}dr\}.
\end{aligned}$$

Here we choose $\Delta_1$, $\Delta_2$ and $\Delta_3$ such that

$$\begin{cases} (1 + \frac{\Delta_1 n}{m})\frac{\partial C}{\partial V} - \frac{\Delta_1}{m} = 0, \\ (1 + \frac{\Delta_1 n}{m})\frac{\partial C}{\partial p} - \Delta_2 \frac{\partial \hat{C}}{\partial p} = 0, \\ (1 + \frac{\Delta_1 n}{m})\frac{\partial C}{\partial r} - \Delta_2 \frac{\partial \hat{C}}{\partial r} - \Delta_3 \frac{\partial Z}{\partial r} = 0 \end{cases},$$

that is,

$$\Delta_1 = m\frac{\partial C}{\partial V}\left(1 - n\frac{\partial C}{\partial V}\right)^{-1}; \quad \Delta_2 = \frac{\partial C}{\partial p}\left(\frac{\partial \hat{C}}{\partial p}\right)^{-1}\left(1 - n\frac{\partial C}{\partial V}\right)^{-1};$$

$$\Delta_3 = \left(1 - n\frac{\partial C}{\partial V}\right)^{-1}\left[\frac{\partial C}{\partial r} - \frac{\partial C}{\partial p}\left(\frac{\partial \hat{C}}{\partial p}\right)^{-1}\frac{\partial \hat{C}}{\partial r}\right]\left(\frac{\partial Z}{\partial r}\right)^{-1}; (1 + \frac{\Delta_1 n}{m}) = \left(1 - n\frac{\partial C}{\partial V}\right)^{-1},$$

(5.10)

then we have

$$d\Pi = \left(1 - n\frac{\partial C}{\partial V}\right)^{-1}\{[\frac{\partial C}{\partial t} + \frac{1}{2}(s_r^2\frac{\partial^2 C}{\partial r^2} + 2\rho_{12}s_r s_V V\frac{\partial^2 C}{\partial r\partial V} + s_V^2 V^2\frac{\partial^2 C}{\partial V^2}$$
$$+ 2\rho_{13}s_r s_p\frac{\partial^2 C}{\partial r\partial p} + s_p^2\frac{\partial^2 C}{\partial p^2})]$$
$$- \frac{\partial C}{\partial p}\left(\frac{\partial \hat{C}}{\partial p}\right)^{-1}[\frac{\partial \hat{C}}{\partial t} + \frac{1}{2}(s_r^2\frac{\partial^2 \hat{C}}{\partial r^2} + 2\rho_{13}s_r s_p\frac{\partial^2 \hat{C}}{\partial r\partial p} + s_p^2\frac{\partial^2 \hat{C}}{\partial p^2})]$$
$$- \left[\frac{\partial C}{\partial r} - \frac{\partial C}{\partial p}\left(\frac{\partial \hat{C}}{\partial p}\right)^{-1}\frac{\partial \hat{C}}{\partial r}\right]\left(\frac{\partial Z}{\partial r}\right)^{-1}(\frac{\partial Z}{\partial t} + \frac{1}{2}s_r^2\frac{\partial^2 Z}{\partial r^2})\}dt.$$

(5.11)

(ii) The case of default (unexpected) in time interval $[t, t + dt]$ (with probability $p_t dt$);
In this case Since we have

$$dC = R_d - C, \quad d\hat{C} = \hat{R} - \hat{C},$$

then substitute them and (5.10) into (5.9), then

$$d\Pi = (1 + \frac{\Delta_1 n}{m})(R_d - C) - \frac{\Delta_1}{m}dV - \Delta_2(\hat{R} - \hat{C}) - \Delta_3 dZ$$

$$= \left(1 - n\frac{\partial C}{\partial V}\right)^{-1}\{(R_d - C) - \frac{\partial C}{\partial V}dV - \frac{\partial C}{\partial p}\left(\frac{\partial \hat{C}}{\partial p}\right)^{-1}(\hat{R} - \hat{C})$$

$$- [\frac{\partial C}{\partial r} - \frac{\partial C}{\partial p}\left(\frac{\partial \hat{C}}{\partial p}\right)^{-1}\frac{\partial \hat{C}}{\partial r}]\left(\frac{\partial Z}{\partial r}\right)^{-1}dZ\}.$$

(5.12)

Thus combine the above two cases (i) and (ii) and then from (5.7)

$$d\Pi = (5.11)\times(\mathbf{1} - \boldsymbol{p_t}\boldsymbol{dt}) + (5.12)\times \boldsymbol{p_t}\boldsymbol{dt} = r\Pi\,dt.$$

Consider (5.8) and (5.10), then

$$\Pi = \left(1 - n\frac{\partial C}{\partial V}\right)^{-1}\left\{C - \frac{\partial C}{\partial V}V - \frac{\partial C}{\partial p}\left(\frac{\partial \hat{C}}{\partial p}\right)^{-1}\hat{C} - [\frac{\partial C}{\partial r} - \frac{\partial C}{\partial p}\left(\frac{\partial \hat{C}}{\partial p}\right)^{-1}\frac{\partial \hat{C}}{\partial r}]\left(\frac{\partial Z}{\partial r}\right)^{-1}Z\right\}.$$

neglect higher order infinitesimal of $\boldsymbol{dt}$ and multiply $1 - n\frac{\partial C}{\partial V}$ to both sides, then we have

$$\frac{\partial C}{\partial t}+\frac{1}{2}(s_r^2\frac{\partial^2 C}{\partial r^2}+2\rho_{12}s_r s_V V\frac{\partial^2 C}{\partial r\partial V}+s_V^2 V^2\frac{\partial^2 C}{\partial V^2}+2\rho_{13}s_r s_p\frac{\partial^2 C}{\partial r\partial p}+s_p^2\frac{\partial^2 C}{\partial p^2})$$

$$+rV\frac{\partial C}{\partial V}-rC+p(R_d-C)$$

$$-\frac{\partial C}{\partial p}\left(\frac{\partial \hat{C}}{\partial p}\right)^{-1}[\frac{\partial \hat{C}}{\partial t}+\frac{1}{2}(s_r^2\frac{\partial^2 \hat{C}}{\partial r^2}+2\rho_{13}s_r s_p\frac{\partial^2 \hat{C}}{\partial r\partial p}+s_p^2\frac{\partial^2 \hat{C}}{\partial p^2})-r\hat{C}+p(\hat{R}-\hat{C})]$$

$$-\left[\frac{\partial C}{\partial r}-\frac{\partial C}{\partial p}\left(\frac{\partial \hat{C}}{\partial p}\right)^{-1}\frac{\partial \hat{C}}{\partial r}\right]\left(\frac{\partial Z}{\partial r}\right)^{-1}(\frac{\partial Z}{\partial t}+\frac{1}{2}s_r^2\frac{\partial^2 Z}{\partial r^2}-rZ)=0.$$

Here, if we consider the equations (4.13) for $\hat{C}(p,t)$ and (4.6) for $Z$, then we have

$$\frac{\partial C}{\partial t}+\frac{1}{2}(s_r^2\frac{\partial^2 C}{\partial r^2}+2\rho_{12}s_r s_V V\frac{\partial^2 C}{\partial r\partial V}+s_V^2 V^2\frac{\partial^2 C}{\partial V^2}+2\rho_{13}s_r s_p\frac{\partial^2 C}{\partial r\partial p}+s_p^2\frac{\partial^2 C}{\partial p^2})$$
$$+\theta(\mu_r-r)\frac{\partial C}{\partial r}+rV\frac{\partial C}{\partial V}+a_p\frac{\partial C}{\partial p}-rC+p(R_d-C)=0. \quad (5.13)$$

*Note* 1: When the assumption (5.2) is not satisfied, using expectation method we can also get (5.13) in risk neutral world.

Thus we have the *PDE model for defaultable bond with Both Expected and Unexpected Defaults in the case with stochastic short rate r and stochastic default intensity p*: Solve the following problem in the region $\Sigma=\{(r,V,p,t):0<t<T,r>0,V>V_B Z,p>0\}$

$$\begin{cases}\frac{\partial C}{\partial t}+\frac{1}{2}(s_r^2\frac{\partial^2 C}{\partial r^2}+2\rho_{12}s_r s_V V\frac{\partial^2 C}{\partial r\partial V}+s_V^2 V^2\frac{\partial^2 C}{\partial V^2}+2\rho_{13}s_r s_p\frac{\partial^2 C}{\partial r\partial p}+s_p^2\frac{\partial^2 C}{\partial p^2})\\ +\theta(\mu_r-r)\frac{\partial C}{\partial r}+rV\frac{\partial C}{\partial V}+a_p\frac{\partial C}{\partial p}-rC+p(RZ(r,t)-C)=0\\ C(r,V,p,T)=1,\\ C(r,V_B Z,p,t)=RZ,\\ C(r\to 0 \text{ or } p\to 0):\text{bounded}\\ C(p\to\infty \text{ or } V\to\infty):\text{bounded}.\end{cases} \quad (5.14)$$

*Remark*: This is a **non-homogeneous** equation with the **coefficients** dependent on **time** and **space variables** and **moving barrier**. In this problem, $\{V=0\}$ are *degenerated boundaries* of $\Sigma$ and we need not to give the boundary value there. In fact, *Fichera's criteria* is

$$B=\sum_{i=1}^{4}\left[b_i(x)-\sum_{j=1}^{4}\frac{\partial a_{ij}(x)}{\partial x_j}\right]\cos(\vec{n},x_i)\Big|_{x\in\{V=0\}},$$

here

$$\boldsymbol{x}=(r,V,p,t), \boldsymbol{b}(\boldsymbol{x})=[b_i]=[\theta(\mu-r), rV, a_p(p,t), 1],$$

$$a_{21}(x)=\rho_{12}s_r s_V V, \quad a_{22}(x)=s_V^2 V^2/2, \quad a_{23}(x)=a_{24}(x)=0$$

Thus

$$B = V[r - s_V^2]\big|_{V=0} = 0 \geq 0.$$

**Solving:**

Use numeraire as zero coupon bond price **Z**, that is, the change of unknown function and variable (5.8)

$$x = \frac{V}{Z}, \quad u(x,p,t) = \frac{C(r,V,p,t)}{Z}. \tag{5.15}$$

Then

$$C_t = u_t Z - xu_x Z_t + uZ_t, \quad VC_V = xu_x Z, \quad C_r = Z_r(u - xu_x), \quad C_p = Zu_p,$$

$$C_{rr} = Z_{rr}(u - xu_x) + x^2 u_{xx} \frac{Z_r^2}{Z}, \quad VC_{Vr} = -x^2 u_{xx} Z_r, \quad V^2 C_{VV} = x^2 u_{xx} Z, \tag{5.16}$$

$$C_{rp} = Z_r(u_p - xu_{xp}), \quad C_{pp} = Zu_{pp}, \quad \frac{Z_r}{Z} = -\overline{B}(t)$$

(Here $\overline{B}(t)$ is defined in (4.7). Substitute (5.15) and (5.16) into (5.14), consider the equation (4.6) for $Z(r, t)$, divide both sides by **Z**, then we get the ***non homogeneous*** equation ***zero risk free rate*** :

$$\begin{cases} u_t + \frac{1}{2}\{[s_r^2 \overline{B}^2(t) + s_V^2 + 2\rho_{12} s_r \overline{B}(t) s_V]x^2 u_{xx} + s_p^2(p,t)u_{pp} \\ \quad + 2\rho_{13} s_r \overline{B}(t) s_p xu_{px}\} + [a_p(p,t) - 2\rho_{13} s_r \overline{B}(t) s_p]u_p \\ \quad - p(u - R) = 0, \quad (x > V_B, \ p > 0, \ 0 < t < T) \\ u(x,p,T) = 1, \quad (x > V_B, \ p > 0) \\ u(V_B, p, t) = R, \quad (p > 0, \ 0 < t < T) \\ u(X \to \infty): \text{bounded}, \quad (p > 0, \ 0 < t < T) \\ u(p \to 0 \text{ or } p \to \infty): \text{bounded} \quad (x > V_b, 0 < t < T) \end{cases} \tag{5.17}$$

Use again the change of unknown function ,

$$\hat{u} = u - R, \tag{5.18}$$

$$\Sigma_x^2(t) = s_r^2 \overline{B}^2(t) + s_V^2 + 2\rho_{12} s_r \overline{B}(t) s_V \geq 0,$$

$$\rho_{p,x}(p,t) = \rho_{13} s_r \overline{B}(t) s_p(p,t), \tag{5.19}$$

$$\overline{a}_p(p,t) = a_p(p,t) - 2\rho_{13} s_r \overline{B}(t) s_p(p,t).$$

then we get the ***homogeneous*** equation with ***risk free rate* 0** ;

$$\begin{cases} \hat{u}_t + \frac{1}{2}[\Sigma_x^2(t)x^2 \hat{u}_{xx} + s_p^2(p,t)\hat{u}_{pp} + 2\rho_{p,x}(p,t)x\hat{u}_{px}] \\ \quad + \overline{a}_p(p,t)\hat{u}_p - p\hat{u} = 0, \quad (x > V_B, \ p > 0, \ 0 < t < T) \\ \hat{u}(x,p,T) = 1 - R, \quad (x > V_B, \ p > 0) \\ \hat{u}(V_B, p, t) = 0, \quad (p > 0, \ 0 < t < T) \\ \hat{u}(X \to \infty): \text{bounded}, \quad (p > 0, \ 0 < t < T) \\ \hat{u}(p \to 0 \text{ or } p \to \infty): \text{bounded} \quad (x > V_b, 0 < t < T) \end{cases} \tag{5.20}$$

Use again the change of unknown function

$$\hat{u} = W(1 - R), \qquad (5.21)$$

then our defaultable bond price is represented by **W** as

$$C(r,V,p,t) = uZ = (\hat{u} + R)Z = [W(1-R) + R]Z$$
$$= W(\frac{V}{Z}, p, t)Z + [1 - W(\frac{V}{Z}, p, t)]RZ \qquad (5.22)$$

That is, the price of our defaultable bond at time *t* can be seen as an expectation of the current value of the bond in the case of no default and the value of the bond in the case of default at time *t*. So *W*(*V/Z*, *p*, *t*) in (5.22) can be regarded as *a probability at time t of no default* in the case of unexpected default intensity is $p_t$ and expected default barrier is $V_B \cdot Z(r, t)$. Thus we get the **PDE equation of no default probability** when default intensity is $p_t$ and expected default barrier is $V_B \cdot Z(r, t)$.

$$\begin{cases} W_t + \frac{1}{2}[\Sigma_x^2(t)x^2 W_{xx} + s_p^2(p,t)W_{pp} + \rho_{p,x}(p,t)xW_{px}] \\ \quad + \overline{a}_p(p,t)W_p - pW = 0, \qquad (x > V_B,\ p > 0,\ 0 < t < T) \\ W(x, p, T) = 1, \qquad (x > V_B,\ p > 0) \\ W(V_B, p, t) = 0, \qquad (p > 0,\ 0 < t < T) \\ W(x \to \infty) : \text{bounded}, \qquad (p > 0,\ 0 < t < T) \\ W(p \to 0 \text{ or } p \to \infty) : \text{bounded} \qquad (x > V_b, 0 < t < T) \end{cases} \qquad (5.23)$$

From the assumption 6) $\rho_{13} = 0$, $\rho_{p,x}(p, t) = 0$ in (5.23), so (5.23) is changed to

$$W_t + \frac{1}{2}[\Sigma_x^2(t)x^2 W_{xx} + s_p^2(p,t)W_{pp}] + a_p(p,t)W_p - pW = 0, \qquad (5.24)$$

and we can guess no default probability can be represented as

$$W(x, p, t) = f(x,t) \cdot g(p,t) \qquad (5.25)$$

Substitute (5.25) into (5.24), then we have

$$\left[ f_t + \frac{1}{2}\Sigma_x^2(t)x^2 f_{xx} \right]g + \left[ g_t + \frac{1}{2}s_p^2(p,t)g_{pp} + a_p(p,t)g_p - pg \right]f = 0.$$

So if we solve the following two problems, then we get the solution of (5.27).

$$\begin{cases} f_t + \frac{1}{2}\Sigma_x^2(t)x^2 f_{xx} = 0, \qquad (0 < t < T,\ x > V_B) \\ f(V_B, t) = 0, \qquad (0 < t < T) \\ f(x, T) = 1, \qquad (V > V_B) \\ f(x \to \infty) : \text{bounded}, \qquad (0 < t < T) \end{cases} \qquad (5.26)$$

$$\begin{cases} g_t + \frac{1}{2}s_p^2(p,t)g_{pp} + a_p(p,t)g_p - pg = 0, \quad (0 < t < T, p > 0) \\ g(p,T) = 1, \qquad (p > 0) \end{cases} \qquad (5.27)$$

*Remark*: Thus, we can regard that *g*(*p*, *t*) is a no default probability with unexpected default intensity $p_t$ and *f*(*V/Z*, *t*) is a no default probability with expected default barrier $V_B Z$.

Under the assumption (5.5), the solution of (5.27) has the form of

$$g(p,t) = e^{A(t,T)-B(t,T)p}, \quad (5.28)$$

here $A(t, T)$ is given by

$$A(t,T) = -\int_t^T [b(s)B(s) - \frac{1}{2}d(s)B^2(s)]ds, \quad (5.29)$$

and in the following special cases we can get $B(t, T)$ as in previous section;

$$B(t,T) = \begin{cases} \dfrac{1-\exp(-c(T-t))}{c}, & dp_t = (b(t) - c \cdot p)dt + \sqrt{d(t)} \cdot dW, c \neq 0, \\ T-t, & dp_t = b(t)dt + \sqrt{d(t)} \cdot dW, \\ \sqrt{2/C}\,\text{th}[\sqrt{C/2}(T-t)], & dp_t = b(t)dt + \sqrt{d(t)+C \cdot p} \cdot dW, C > 0. \end{cases} \quad (5.30)$$

(5.26) can be thought as a pricing equation of a "**bond with barrier** with *zero risk free rate*, *zero dividend rate* and time dependent volatility".

$$s = \int_0^t \Sigma_x^2(u)du, \quad T^* = \int_0^T \Sigma_x^2(u)du, \quad f(x,t) = \bar{f}(x,s)! \quad (5.31)$$

$$\begin{cases} \bar{f}_s + \dfrac{1}{2}x^2 \bar{f}_{xx} = 0, & (0 < s < T^*, \ x > V_B) \\ \bar{f}(V_B, s) = 0, & (0 < s < T^*) \\ \bar{f}(x, T^*) = 1, & (V > V_B) \\ \bar{f}(x \to \infty): \text{bounded}, & (0 < t < T) \end{cases} \quad (5.32)$$

Using the same method used in *put-call parity* for down and out barrier options (see ch. 8, theorem **1** of [Jiang LS 03]), we can get

$$\bar{f}(x,s) = \left[ N(d_1) - \left(\dfrac{x}{V_B}\right) N(d_2) \right], \quad (5.33)$$

$$d_1 = \dfrac{\ln\dfrac{x}{V_B} - \dfrac{1}{2}(T^* - s)}{\sqrt{T^* - s}}, \quad d_2 = \dfrac{\ln\dfrac{V_B}{x} - \dfrac{1}{2}(T^* - s)}{\sqrt{T^* - s}} \quad (5.34)$$

Returning to original variable, then

$$f(x,t) = \left[ N(\bar{d}_1) - \left(\dfrac{x}{V_B}\right) N(\bar{d}_2) \right], \quad (5.35)$$

$$\bar{d}_1 = \dfrac{\ln\dfrac{x}{V_B} - \dfrac{1}{2}\int_t^T \Sigma_x^2(u)du}{\sqrt{\int_t^T \Sigma_x^2(u)du}} = \dfrac{\ln\dfrac{V}{V_B Z} - \dfrac{1}{2}\int_t^T \Sigma_x^2(u)du}{\sqrt{\int_t^T \Sigma_x^2(u)du}},$$

$$\bar{d}_2 = \frac{\ln\frac{V_B}{x} - \frac{1}{2}\int_t^T \Sigma_x^2(u)du}{\sqrt{\int_t^T \Sigma_x^2(u)du}} = \frac{\ln\frac{V_B Z}{V} - \frac{1}{2}\int_t^T \Sigma_x^2(u)du}{\sqrt{\int_t^T \Sigma_x^2(u)du}}. \qquad (5.36)$$

Here $\Sigma_x^2(u)$ is given by (5.19), (4.7).

*The No default probability formula*:

$$W\left(\frac{V}{Z}, p, t\right) = e^{A(t,T) - B(t,T)p}\left[N(\bar{d}_1) - \left(\frac{V}{V_B Z}\right)N(\bar{d}_2)\right], \qquad (5.37)$$

*The Pricing formula* of *defaultable zero coupon corporate bond* with **stochastic unexpected default intensity** and expected **default barrier**:

$$\begin{aligned}C(r,V,p,t) &= RZ + W(\frac{V}{Z}, p, t)(1-R)Z \\ &= RZ + e^{A(t,T) - B(t,T)p}(1-R)\left[ZN(\bar{d}_1) - \left(\frac{V}{V_B}\right)N(\bar{d}_2)\right],\end{aligned} \qquad (5.38)$$

$\bar{d}_1, \bar{d}_2$ in (5.37) and (5.38) are defined in (5.36).

The *financial meaning* of (5.38) is clear: the *first term* of (5.38) is the current price of *the part to be given to bond holder **regardless of default occurs or not***, and the *second term* is the allowance dependent on default probability at time *t*. If at some moment *t*, the default is certain ($W(V/Z, p, t) = 0$), then the price of the bond at *t* is exactly the current price of default recovery. And as shown in formula (5.38), if $R = 0$, that is, if there is nothing to recover in case of default, then the bond price is the product of zero coupon bond and survival probability; and if $R = 1$, that is, if there is no any loss in case of default, then the bond price is the same with zero coupon bond.

Using the same method, we can derive a *price formula of a credit default swap*. Consider a credit default swap which pays $(1-R) \cdot Z$ in case of unexpected default and the expected default before time *T* to the corporate bond holder and it require the protection buyer must pay the price of the credit default swap at the first date of that insurance contract. Let denote our swap by $S(Z, V, p, t)$, then it satisfies the following problem in the region

$$\Sigma = \{(r, V, p, t) : 0 < t < T, r > 0, V > V_B Z, p > 0\}$$

$$\begin{cases} \frac{\partial S}{\partial t} + \frac{1}{2}(s_r^2 \frac{\partial^2 S}{\partial r^2} + 2\rho_{12} s_r s_V V \frac{\partial^2 S}{\partial r \partial V} + s_V^2 V^2 \frac{\partial^2 S}{\partial V^2} + 2\rho_{13} s_r s_p \frac{\partial^2 S}{\partial r \partial p} + s_p^2 \frac{\partial^2 S}{\partial p^2}) \\ \quad + \theta(\mu_r - r)\frac{\partial S}{\partial r} + rV\frac{\partial S}{\partial V} + a_p \frac{\partial S}{\partial p} - (r+p)S + p(1-R)Z(r,t) = 0 \\ S(r, V, p, T) = 0, \\ S(r, V_B Z, p, t) = (1-R)Z, \\ S(r \to 0 \text{ or } p \to 0) : \text{bounded} \\ S(p \to \infty \text{ or } V \to \infty) : \text{bounded}. \end{cases} \qquad (5.39)$$

Using the same method as the above, we can get

$$S(r,V,p,t) = (1-W)(1-R)Z, \quad (5.40)$$

Here $W$ is the no default probability given by (5.37).

The **financial meaning** of (5.40) is clear: since the swap contract gives nothing in case of no default and gives compensation $(1-R) \cdot Z$ in case of default, then the price of the swap contract is the product of the compensation and default probability.

**Remark 1.** In this section, the assumption 3) about expected default barrier, the assumption 5) about default recovery function and the assumption 6) about no correlations of unexpected default intensity $p_t$ with firm value $V$ and short rate $r$ made an important role in deriving analytical pricing formula. If the default recovery value was not a discounted value but a constant, then it is rather difficult to change non homogeneous equation (5.14) into homogeneous equation (5.20). And if expected default recovery and unexpected default recovery were not same, then our logic would not work and it might need another method.

**Remark 2.** In this section, the assumption (5.1) that the firm value is sum of some shares of stocks and defaultable bonds made an important role in hedged the risk of firm value $V$ In this section, we hedged the risk of firm value $V$ using (5.1), the risk of unexpected default intensity $p$ using the "virtual bond" with only unexpected default and the risk of short rate $r$ using default free zero coupon bond $Z$, so our equation (5.13) has no any drifts on natural measure and it gives fair price to our bonds.

**Remark 3.** The assumption (5.2) about the model of short rate is not essential in deriving pricing formula but the fact that the default free zero coupon bond price $Z$ is given by exponential function of type (4.7) are essential for our formula. So not only Vasicek Model but **Ho-Lee model** or **Hull-White model** also do work well.

**Remark 4.** In the case when the drift and squared volatility of default intensity $p_t$ are linear on $r$ and $p$ but do not satisfy the special conditions like (4.32), we have **semi-analytical formula** of bond price, because $C(t, T)$ in pricing formulae cannot have an explicit formula.

## $6. Credit Spread Analysis

In this section, we will illustrate properties of credit spreads implied by the above model. For simplicity, we deal with constant short rate case. The credit spread is defined as the difference between the yields of defaultable and default-free bonds and is given by the following expression:

$$CS = -\frac{\log(C(V,p,t)/e^{-r(T-t)})}{T-t} \quad (4.1)$$

This simplifies in our model to

$$CS = -\frac{\log(R + W(V,p,t)(1-R))}{T-t} \quad (4.2)$$

In order to study the properties of credit spreads, we consider a *base case* environment with the following parameters values:

current time $t = 0$, default recovery $R = 0.5$, short rate $r = 0.07$, firm dividend rate $b = 0.03$,

firm value volatility $s_V = 0.2$, $V/V_B = 1.5$;

and in unexpected default intensity model (2.8) and (2.16) or (3.23), we set

$$b(t) = 0.1, c(t) = 1.4248 \times 0.0038, d(t) = 0.0131^2, e(t) = 0.$$

(Collin-Solnik (2001) obtained maximum likelihood estimates of the credit spread parameters using investment grade corporate bonds data and we followed their estimates as Hou (2003) did).

In the following, we will show the effects of default intensity on the magnitude and shape of the term structure of credit spreads. Figure 1 and 4 show, as one would expect that an increase in default intensity increases the credit spreads. Figure 2 and 5 show the effects of default intensity on the term structure of credit spreads. The credit spreads derived are different from zero for very short maturity and the effect decrease as maturity increase. Figure 3 and 6 show the effects of default intensity on the term structure of bond price and the effect increase as maturity increase.

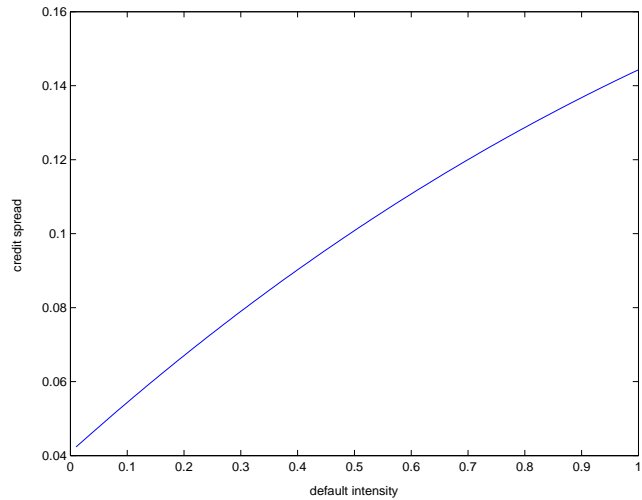

**Figure 1**. credit spread – default intensity plot ($CS$ - $p$), $p = 0.1 \sim 1$, $V_b(t) = V_B$

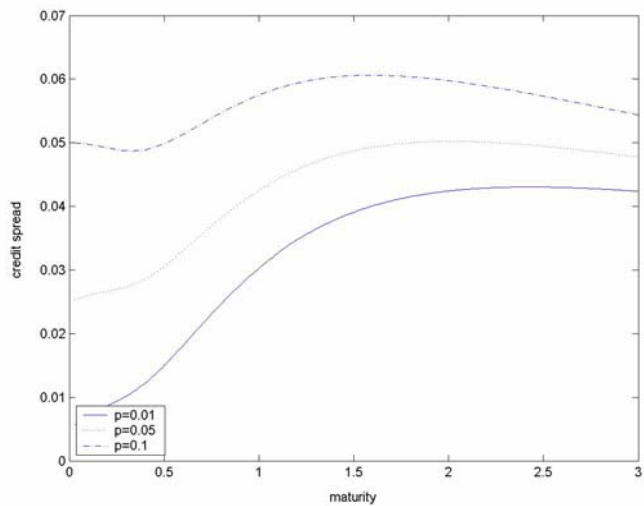

**Figure 2**. credit spread – maturity plot ($CS$ - $T$), $T = 0.1 \sim 3$, $V_b(t) = V_B$

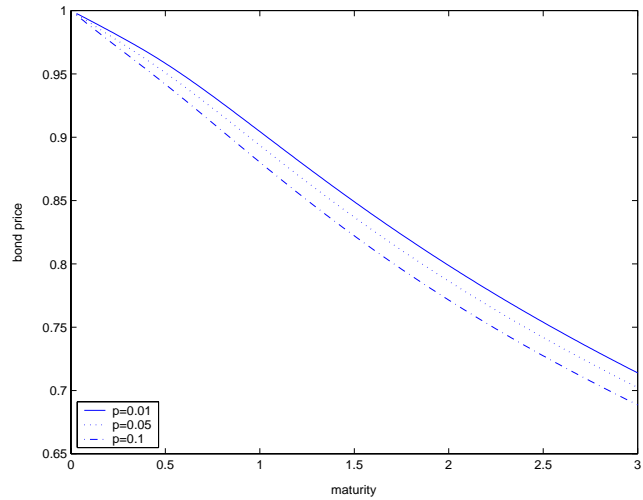

**Figure 3**. bond price – maturity plot ($C$ - $T$), $T = 0.1 \sim 3$, $V_b(t) = V_B$

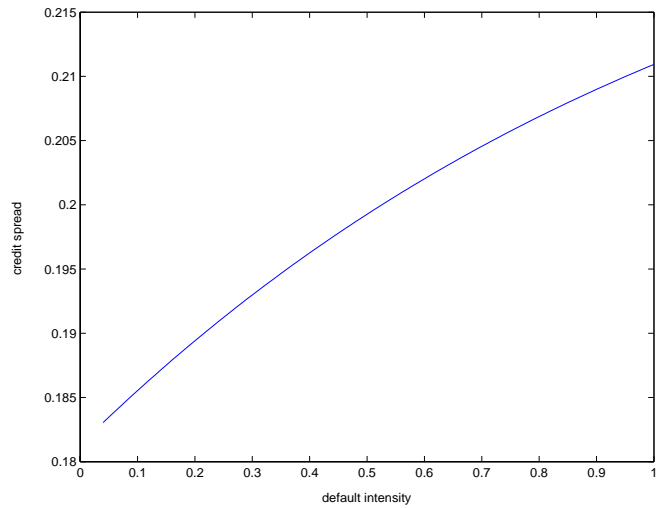

**Figure 4**. credit spread – default intensity plot ($CS$ - $p$), $p = 0.1 \sim 1$, $V_b(t) = V_B e^{-r(T-t)}$

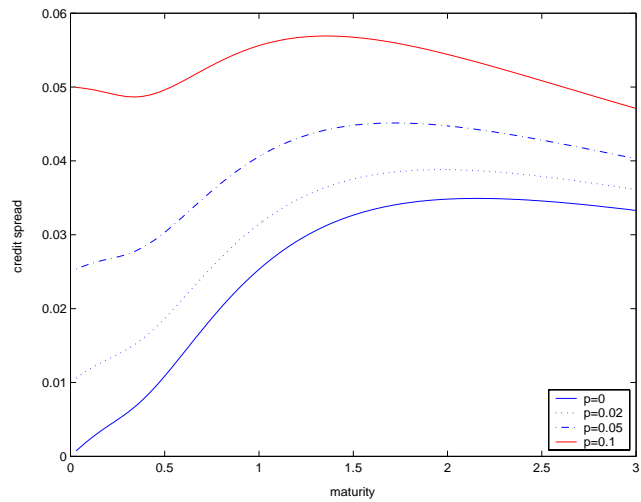

**Figure 5**. credit spread – maturity plot ($CS$ - $T$), $T = 0.1 \sim 3$, $V_b(t) = V_B e^{-r(T-t)}$

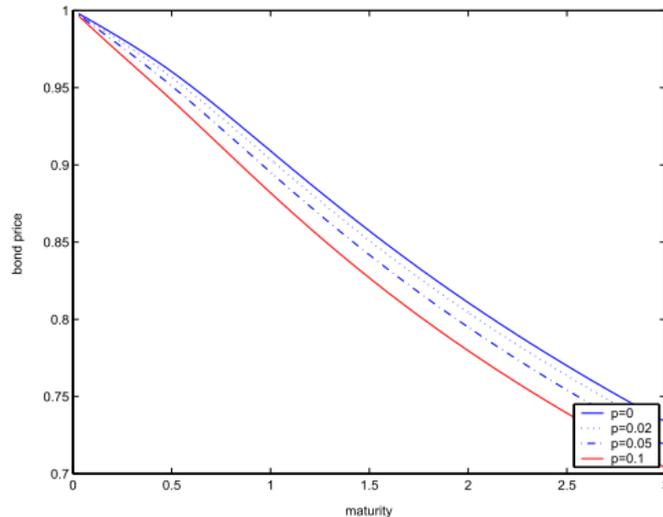

**Figure 6**. bond price – maturity plot (*C* - *T*), $T = 0.1 \sim 3$, $V_b(t) = V_B e^{-r(T-t)}$

**Acknowledgement:** Authors would like to thank Professors Li-shang Jiang and Bao-jun Bian for their useful advices and information.

## Appendix 1. The Proof of (2.18) and (2.19)

From (2.17) we have

$$W_t = W(A' - B'p); \quad W_p = -WB; \quad W_{pp} = WB^2, \tag{A.1}$$

and substitute (A.1), (2.17) and (2.16) into (2.14) and divide by $W$, then

$$A' + \frac{1}{2}d(t)B^2 - b(t)B + p(-B' + \frac{1}{2}e(t)B^2 + c(t)B - 1) = 0 \tag{A.2}$$

Differentiate on $p$, then

$$B' - c(t)B - \frac{1}{2}e(t)B^2 + 1 = 0. \tag{A.3}$$

So if we consider (A.3) in (A.2), then

$$A' + \frac{1}{2}d(t)B^2 - b(t)B = 0 \tag{A.4}$$

From $W(p, T) = 1$, we know that

$$A(T, T) = B(T, T) = 0$$

(A.3) is well known **Riccati equation** in the case of $e(t) \neq 0$. If $B$ would be known, then (A.4) is easily integrated:

$$A(t,T) = -\int_t^T [b(s)B(s) - \frac{1}{2}d(s)B^2(s)]ds \tag{A.5}$$

where we took into account $A(T, T) = 0$.

Generally, (A.3) can not be solved, but *in many important cases, we can solve it*.

**Case 1)** $c(t) \equiv c$ (const); $e(t) \equiv 0$; that is

$$dp_t = (b(t) - c \cdot p)dt + \sqrt{d(t)} \cdot dW \tag{A.6}$$

(This case includes *Vasicek model* (b, c, d : const, e = 0), *Ho-Lee model* (c = 0, e = 0, d: const) and *Hull-White model* (c, d: const, e = 0) ). Then (A.3) is

$$B' = cB - 1,$$

the solution is

$$B(t,T) = \begin{cases} \dfrac{1-\exp(-c(T-t))}{c}, & c \neq 0, \\ T-t, & c = 0. \end{cases} \tag{A.7}$$

**Case 2)** $c(t) \equiv 0$; $e(t) \equiv C > 0$ (const); that is

$$dp_t = b(t)dt + \sqrt{d(t) + C \cdot p} \cdot dW \tag{A.8}$$

Then (A.3) is

$$B' - \frac{C}{2}B^2 + 1 = 0,$$

the solution is given as follows [Kamke 1980]:

$$B(t,T) = \sqrt{2/C}\, \text{th}[\sqrt{C/2}(T-t)], \qquad (A.9)$$

where

$$\text{th}(x) = \frac{e^x - e^{-x}}{e^x + e^{-x}}. \qquad (A.10)$$

## Appendix 2. The Proof of (4.32), (4.33) and (4.34)

From (4.31)

$$\begin{aligned}
&\hat{C}_t = \hat{C}\cdot(A'(t) - B'(t)r - C'(t)p); \\
&\hat{C}_r = -\hat{C}\cdot B(t); \qquad \hat{C}_{rr} = \hat{C}\cdot B^2(t); \\
&\hat{C}_p = -\hat{C}\cdot C(t); \qquad \hat{C}_{pp} = \hat{C}\cdot C^2(t)
\end{aligned} \qquad (A.11)$$

and substitute (4.31), (A.11) and (4.3) into (4.30), then

$$\begin{aligned}
\hat{C}\cdot\{&[A'(t) + \frac{1}{2}s_r^2\cdot B^2(t) + \frac{1}{2}\delta(t)C^2(t) - \theta\mu_r B(t) - \alpha(t)C(t)] \\
&+ r[-B'(t) + \frac{1}{2}\varepsilon(t)C^2(t) + \theta B(t) - \beta(t)C(t) - 1] \\
&+ p[-C'(t) + \frac{1}{2}\phi(t)C^2(t) - \chi(t)C(t)) - 1 + R]\} = 0.
\end{aligned}$$

Divide both sides by *C* and differentiate on *p* and *r* then we get

$$A'(t) = \theta\mu_r B(t) + \alpha(t)C(t) - \frac{1}{2}s_r^2\cdot B^2(t) - \frac{1}{2}\delta(t)C^2(t), \qquad (A.12)$$

$$B'(t) - \theta B(t) = \frac{1}{2}\varepsilon(t)C^2(t) - \beta(t)C(t) - 1, \qquad (A.13)$$

$$C'(t) - \frac{1}{2}\phi(t)C^2(t) + \chi(t)C(t) = R - 1, \qquad (A.14)$$

$$A(T,T) = B(T,T) = C(T,T) = 0. \qquad (A.15)$$

Once (A.14) is solved, then (A.12) and (A.13) are easily solved. (A.14) is well-known **Riccati equation** in the case of $\chi(t) \neq 0$. Generally, (A.14) can not be solved, but *in many important cases, we can solve it.*

**Case 1)** $\chi(t) \equiv c$ (const); $\phi(t) \equiv 0$; that is

$$a_p(r,p,t) = \alpha(t) + \beta(t)r + c\cdot p, \quad s_p^2(r,p,t) = \delta(t) + \varepsilon(t)r, \qquad (A.16)$$

or

$$dp = [\alpha(t) + \beta(t)r + c\cdot p]dt + \sqrt{\delta(t) + \varepsilon(t)r}\cdot dW, \qquad (A.17)$$

then (A.14) becomes

$$C'(t) = -c\cdot C(t) + R - 1,$$

Analytical Pricing of Defaultable Bond with Stochastic Default Intensity–Exogenous Default Recoverythe solution is

$$C(t,T) = \begin{cases} \dfrac{R-1+(1-R)\exp(c(T-t))}{c}, & c \neq 0, \\ (1-R)(T-t), & c = 0. \end{cases} \quad (A.18)$$

**Case 2)** $\chi(t) \equiv 0; \quad \phi(t) \equiv c > 0$ (const); that is

$$a_p(r,p,t) = \alpha(t) + \beta(t)r, \quad s_p^2(r,p,t) = \delta(t) + \varepsilon(t)r + c \cdot p, \quad (A.19)$$

or

$$dp = [\alpha(t) + \beta(t)r]dt + \sqrt{\delta(t) + \varepsilon(t)r + c \cdot p} \cdot dW, \quad (A.10)$$

Then (A.14) becomes

$$C'(t) = \frac{c}{2}C^2(t) + R - 1,$$

the solution is given as follows [Kamke 1980, p342]:

$$C(t,T) = \sqrt{\frac{2(1-R)}{c}} \text{th}\left[\sqrt{\frac{(1-R)c}{2}}(T-t)\right], \quad (A.21)$$

35